# Stationary solutions of second-order equations for fermions in Reissner-Nordström space-time

V.P.Neznamov[1,2*], I. I. Safronov[1], V.E. Shemarulin[1]

[1]FSUE "RFNC-VNIIEF", Russia, Sarov, Mira pr., 37, 607188
[2]National Research Nuclear University MEPhI, Moscow, Russia

Abstract

Existence of degenerate stationary bound states with square integrable radial wave functions was proved when second-order equations are used with the effective potential of the Reissner-Nordström (RN) field with two event horizons for charged and uncharged fermions,. The fermions in such states are localized near event horizons within the ranges from zero to several fractions of Compton wave length of fermions versus the values of gravitational and electromagnetic coupling constants and the values of angular and orbital momenta $j,l$.

In case of extreme RN fields, absence of stationary bound states of fermions with the energies of $E < mc^2$ is shown for solutions of the second-order equation for any value of gravitational and electromagnetic coupling constants.

Existence of the discrete energy spectrum is shown for the naked RN singularity due to solution of the second-order equation at definite values of physical parameters. The discrete spectrum exists for both charged and uncharged fermions. The naked RN singularity in quantum mechanics with the second-order equation for half-spin particles poses no threat to cosmic censorship since it is covered with an infinitely large potential barrier.

Electrically neutral systems of atomic type (RN collapsars with the definite number of fermions in degenerate bound states) are proposed to consider as particles of dark matter.

*Keywords: Reissner-Nordström metric, self-conjugate Hamiltonian, Dirac equations, Scrödinger-type equations with effective potential, Prüfer transformation, stationary bound states, cosmic censorship, dark matter*

---

[*] vpneznamov@vniief.ru
vpneznamov@mail.ru

## 1. Introduction

Earlier in [1], existence of the degenerate stationary bound state of fermions with binding energy $E_b = mc^2$ (where $m$ is fermion mass, $c$ is light velocity) was proved for the Schwarzschild metric when using self-conjugate second-order equation with the spinor wave function. Energies of degenerate stationary bound states of fermions were announced ibidem for Reissner-Nordström, Kerr, Kerr-Newman fields.

In this paper, we discuss stationary solutions of the second-order equation for fermions in the RN field. Initially, we specify one more time the absence of stationary regular solutions of the Dirac equation in the RN field. The causes of that are:

- square nonintegrability of wave functions in the neighborhood of event horizons (see also [2]);

- non-fulfillment of one of the conditions of stationary bound state existence of fermions in the extreme RN field in the domain outside the single event horizon proved earlier in [3];

- existence of two regular solutions with square integrable wave functions [4] for the domain under the event horizon of the extreme RN field and for the case of naked RN singularity.

The situation qualitatively changes, when using the self-conjugate second-order equation with the spinor wave function for fermions:

- in the paper, existence of degenerate stationary bound states with appropriate square integrable radial wave functions is validated and proved for the RN metric with two event horizons for charged and uncharged fermions;

- conversely, impossibility of existence of bound states is proved for fermions outside the event horizon for the extreme RN field with the single event horizon. The bound state is possible under the event horizon in case of like charges of the fermion and the RN field source for the fermion energy $E > mc^2$ [5];

- existence of the discrete spectrum of half-spin particles is validated and numerically shown for naked RN singularity at definite values of physical parameters;

- it is demonstrated in the paper that the field of naked RN singularity in quantum mechanics with the second-order equation for fermions poses no threat to the cosmic censorship, since the singularity is covered with the infinitely large potential barrier. This conclusion confirms the results of [6], obtained in quantum mechanics of spinless particles for a number of time-like naked singularities (for the Reissner-Nordström metric included).

The paper is arranged as follows. In section 2, the self-conjugate Dirac Hamiltonian is obtained in the RN field with the plane scalar product of wave functions, the variables are separated, the asymptotic form of radial wave functions is explored, the existence of non-regular stationary solutions is found out. In case of asymptotics at $r \to 0$, existence of two square integrable solutions of Dirac equation is found out according to [4].

Section 3 is dedicated to the self-conjugate second-order equation with the spinor wave function. The wave functions of the equation are square integrable in the neighborhood of event horizons and allow one to specify a physically acceptable choice of one of two regular solutions at $r \to 0$. The singularities of effective potentials testify to the possibility of existence of bound states of fermions both in case of existence of two event horizons and in case of RN naked singularity. Conversely, in this case, absence of the potential well for the extreme RN field testifies to the absence of bound states of fermions $E < mc^2$. In section 4, the second-order equation with the effective potential after Prüfer transformation [7] - [10] and introduction of the phase function is presented as a system of first-order non-linear differential equations.

In section 5, the results are presented for numerical calculations with discussion of their performance procedure in the neighborhood of singular points $r = \infty$, $r = 0$ and $r = r_{\pm}$, where $r_{\pm}$ are radii of external and internal event horizons of the RN field. In section 6, absence of threat to the cosmic censorship is found out in quantum mechanics of fermion motion in the RN field of naked singularity. In section 7, electrically neutral atomic systems with fermions in degenerate bound states are proposed as particles of dark matter. In Conclusions, basic results are presented and discussed. Appendixes A, B, present the procedure for obtaining and the explicit form of effective potentials of the second-order equation in the RN field.

## 2. Dirac equation in the Reissner-Nordström field

In the paper, we will generally use the system of units $\hbar = c = 1$, the signature of the metric of plane space-time is chosen to be

$$\eta_{\underline{\alpha\beta}} = \text{diag}[1,-1,-1,-1]. \tag{1}$$

In (1) and below, the underlined indexes are local.

The indexes with Greek letters assume the values of $0, 1, 2, 3$; the indexes with Latin letters assume the values of $1, 2, 3$. The standard rule is used for summation over repeated indexes.

### 2.1 Reissner-Nordström metric

The static RN metric is characterized by a point source of mass $M$ and charge $Q$

$$ds^2 = f_{R-N}dt^2 - \frac{dr^2}{f_{R-N}} - r^2\left(d\theta^2 + \sin^2\theta d\varphi^2\right). \tag{2}$$

In (2), $g_{00} = f_{R-N}$, $g^{00} = (f_{R-N})^{-1}$, $f_{R-N} = \left(1 - \frac{r_0}{r} + \frac{r_Q^2}{r^2}\right)$, $r_0 = 2GM/c^2$ is the gravitational radius of the Schwarzschild field, $r_Q = \sqrt{G}Q/c^2$, $G$ is the gravitational constant, $c$ is the light velocity.

1. If $r_0^2 > 4r_Q^2$,

$$f_{R-N} = \left(1 - \frac{r_+}{r}\right)\left(1 - \frac{r_-}{r}\right), \tag{3}$$

where $r_\pm$ are radii of external and internal event horizons

$$r_\pm = \frac{r_0}{2} \pm \sqrt{\frac{r_0^2}{4} - r_Q^2}. \tag{4}$$

2. The case of $r_0^2 = 4r_Q^2$ corresponds to the extreme RN field with the single event horizon $r_\pm = r_0/2$

3. The case of $r_0^2 < 4r_Q^2$ corresponds to the naked singularity. In this case, $f_{R-N} > 0$.

**2.2 Self-conjugate Hamiltonian**

The algorithms for obtaining self-conjugate Dirac Hamiltonians in external gravitational fields by methods of pseudo-Hermitian quantum mechanics are presented in [11] - [13].

The Dirac equations in the Hamiltonian form for the half-spin particle of mass $m$ and charge $q$ in the RN field have the form of

$$i\frac{\partial \Psi_\eta(\mathbf{r},t)}{\partial t} = H_\eta \Psi_\eta(\mathbf{r},t), \tag{5}$$

$H_\eta = H_\eta^+$ is the self-conjugate Hamiltonian with the plane scalar product of the wave functions.

For diagonal metric tensors $g_{\mu\nu}$, the Hamiltonian $H_\eta$ is easy to derive from the equality obtained in [13]:

$$H_\eta = \left(\tilde{H}_{red} + \tilde{H}_{red}^+\right)/2, \tag{6}$$

$$\tilde{H}_{red} = \frac{m}{g^{00}}\tilde{\gamma}^0 - \frac{i}{g^{00}}\tilde{\gamma}^0\tilde{\gamma}^k\frac{\partial}{\partial x^k} + qA^0. \tag{7}$$

In (7), summation over $k = 1, 2, 3$ is implied.

In equalities (5) - (7), the following denotations are taken. The sign "+" stands for Hermitian conjugacy. The sign "~" above the values means that they have been obtained by using tetrad vectors in Schwinger gauge [14]. $A^0 = Q/r$ is the scalar electromagnetic potential for the RN metric. $\tilde{\gamma}^\mu (\mu = 0,1,2,3)$ are Dirac matrices with world indices. The values of $\tilde{\gamma}^\mu$ relate to matrices of $\gamma^{\underline{\beta}}$ through tetrad vectors in Schwinger gauge $\left(\tilde{\gamma}^\mu = \tilde{H}^\mu_{\underline{\beta}} \gamma^{\underline{\beta}}\right)$. The non-zero tetrad vectors in Schwinger gauge for the RN metric are

$$H^0_{\underline{0}} = \left(\sqrt{f_{R-N}}\right)^{-1}; \quad H^1_{\underline{1}} = \sqrt{f_{R-N}}; \quad H^2_{\underline{2}} = r^{-1}; \quad H^3_{\underline{3}} = (r\sin\theta)^{-1}. \tag{8}$$

Taking into account (6) - (8), the following expression can be obtained for the self-conjugate Hamiltonian $H_\eta = H^+_\eta$ in spherical coordinates $(r, \theta, \varphi)$:

$$\begin{aligned}
H_\eta &= \sqrt{f_{R-N}} m \gamma^{\underline{0}} - i \gamma^{\underline{0}} \gamma^{\underline{1}} \left( f_{R-N} \frac{\partial}{\partial r} + \frac{1}{r} - \frac{r_0}{2r^2} \right) - \\
&\quad - i \sqrt{f_{R-N}} \frac{1}{r} \left[ \gamma^{\underline{0}} \gamma^{\underline{2}} \left( \frac{\partial}{\partial \theta} + \frac{1}{2} \text{ctg}\,\theta \right) + \gamma^{\underline{0}} \gamma^{\underline{3}} \frac{1}{\sin\theta} \frac{\partial}{\partial \varphi} \right] + \frac{qQ}{r}.
\end{aligned} \tag{9}$$

In (9), $\gamma^{\underline{0}}, \gamma^{\underline{k}}$ are Dirac matrices with local indexes.

The expression in Hamiltonian (9), contained in square brackets, depends only on angular coordinates, the rest of the summands depend on the radial coordinate alone.

### 2.3 Separation of variables

For separation of variables, let us present the bispinor $\Psi_\eta(\mathbf{r}, t)$ as

$$\Psi_\eta(r, \theta, \varphi, t) = \begin{pmatrix} F(r) \xi(\theta) \\ -iG(r) \sigma^3 \xi(\theta) \end{pmatrix} e^{im_\varphi \varphi} e^{-iEt} \tag{10}$$

and use the Brill –Wheeler equation [15]

$$\left[ -\sigma^2 \left( \frac{\partial}{\partial \theta} + \frac{1}{2} \text{ctg}\,\theta \right) + i\sigma^1 m_\varphi \frac{1}{\sin\theta} \right] \xi(\theta) = i\kappa \xi(\theta). \tag{11}$$

In order to use the equation, it is convenient to perform the equivalent substitution of matrices in Hamiltonian (9):

$$\gamma^{\underline{1}} \to \gamma^{\underline{3}}, \quad \gamma^{\underline{3}} \to \gamma^{\underline{2}}, \quad \gamma^{\underline{2}} \to \gamma^{\underline{1}} \tag{12}$$

In equalities (10), (11): $\xi(\theta)$ are spherical harmonics for a half-spin, $\sigma^k$ are two-dimensional Pauli matrices, $E$ is the energy of the Dirac particle, $m_\varphi = -j, -j+1, \ldots j$ is the azimuthal component of angular momentum $j$, $\kappa$ is the quantum number of the Dirac equation:

$$\kappa = \mp 1, \mp 2, \ldots = \begin{cases} -(l+1), & j = l + 1/2, \\ l, & j = l - 1/2, \end{cases} \tag{13}$$

$j, l$ are quantum numbers of the total angular and orbital momenta of Dirac particles. $\xi(\theta)$ can be presented as [16]

$$\xi(\theta) = \begin{pmatrix} _{-1/2}Y_{jm_\varphi}(\theta) \\ _{1/2}Y_{jm_\varphi}(\theta) \end{pmatrix} = (-1)^{m_\varphi + 1/2} \sqrt{\frac{1}{4\pi}\frac{(j-m_\varphi)!}{(j+m_\varphi)!}} \begin{pmatrix} \cos\theta/2 & \sin\theta/2 \\ -\sin\theta/2 & \cos\theta/2 \end{pmatrix} \times \\ \times \begin{pmatrix} (\kappa - m_\varphi + 1/2) P_l^{m_\varphi - 1/2}(\theta) \\ P_l^{m_\varphi + 1/2}(\theta) \end{pmatrix}. \tag{14}$$

In (14), the expression after the square root in the parentheses is a two-dimensional matrix, $P_l^{m_\varphi \pm 1/2}(\theta)$ are associated Legendre polynomials.

As the result of separation of variables, we obtain the equations for the radial functions of $F(\rho), G(\rho)$:

$$f_{R-N}\frac{dF(\rho)}{d\rho} + \left(\frac{1+\kappa\sqrt{f_{R-N}}}{\rho} - \frac{\alpha}{\rho^2}\right)F(\rho) - \left(\varepsilon - \frac{\alpha_{em}}{\rho} + \sqrt{f_{R-N}}\right)G(\rho) = 0,$$

$$f_{R-N}\frac{dG(\rho)}{d\rho} + \left(\frac{1-\kappa\sqrt{f_{R-N}}}{\rho} - \frac{\alpha}{\rho^2}\right)G(\rho) + \left(\varepsilon - \frac{\alpha_{em}}{\rho} - \sqrt{f_{R-N}}\right)F(\rho) = 0. \tag{15}$$

In (15), the dimensionless variables are introduced:

$$\rho = \frac{r}{l_c}; \quad \varepsilon = \frac{E}{mc^2}; \quad \alpha = \frac{r_0}{2l_c} = \frac{GMm}{\hbar c} = \frac{Mm}{M_P^2},$$

$$\alpha_Q = \frac{r_Q}{l_c} = \frac{\sqrt{G}Qm}{\hbar c} = \frac{\sqrt{\alpha_{fs}}}{M_P}m\frac{Q}{e}; \quad \alpha_{em} = \frac{qQ}{\hbar c} = \alpha_{fs}\frac{q}{e}\frac{Q}{e}. \tag{16}$$

Here, $l_c = \hbar/mc$ is the Compton wave length of a Dirac particle; $M_P = \sqrt{\hbar c/G} = 2.2 \cdot 10^{-5}$ g $(1.2 \cdot 10^{19} GeV)$ is the Planck mass; $\alpha_{fs} = e^2/\hbar c \approx 1/137$ is the electromagnetic constant of the fine structure; $\alpha, \alpha_{em}$ are gravitational and electromagnetic coupling constants; $\alpha_Q$ is the dimensionless constant, characterizing the source of the electromagnetic field in the RN metric.

In denotations of (16),

$$f_{R-N} = 1 - \frac{2\alpha}{\rho} + \frac{\alpha_Q^2}{\rho^2}. \tag{17}$$

In case of availability of external and internal event horizons: $\alpha^2 > \alpha_Q^2$ and

$$f_{R-N} = \frac{(\rho - \rho_+)(\rho - \rho_-)}{\rho^2}, \tag{18}$$

where

$$\rho_{\pm} = \alpha \pm \sqrt{\alpha^2 - \alpha_Q^2}. \tag{19}$$

For the extreme RN field: $\alpha^2 = \alpha_Q^2$, $\rho_+ = \rho_- = \alpha$ and

$$f_{R-N} = \frac{(\rho - \alpha)^2}{\rho^2}. \tag{20}$$

The case of naked RN singularity is implemented at $\alpha^2 < \alpha_Q^2$.

**2.4 Asymptotics of radial wave functions**

**2.4.1 Presence of event horizons $\rho_+, \rho_-$ $\left(\alpha^2 > \alpha_Q^2\right)$**

At $\rho \to \infty$, the leading terms of the asymptotics are (see, for instance, [17], [18])

$$F = C_1 \varphi_1(\rho) e^{-\sqrt{1-\varepsilon^2}\rho} + C_2 \varphi_2(\rho) e^{\sqrt{1-\varepsilon^2}\rho},$$
$$G = \sqrt{\frac{1-\varepsilon}{1+\varepsilon}} \left( -C_1 \varphi_1(\rho) e^{-\sqrt{1-\varepsilon^2}\rho} + C_2 \varphi_2(\rho) e^{\sqrt{1-\varepsilon^2}\rho} \right). \tag{21}$$

In (21) $\varphi_1(\rho), \varphi_2(\rho)$ are power functions of $\rho$. To ensure the finite motion of Dirac particles, it is necessary to use only exponentially decreasing solutions (21), i.e., in this case, $C_2 = 0$.

At $\rho \to 0$ in [4], existence of two square integrable solutions of the Dirac equation with the roots of the indicial equation $s_1 = 0$, $s_2 = 2$ was proved for functions $f(\rho) = \rho F(\rho)$, $g(\rho) = \rho G(\rho)$. In this case, it is impossible to formulate the boundary problem for obtaining stationary solutions of the Dirac equation corresponding to bound states of fermions.

At $\rho \to \rho_+$, let us present the functions $F(\rho), G(\rho)$ as

$$F\big|_{\rho \to \rho_+} = \left(|\rho - \rho_+|\right)^{s_+} \sum_{k=0}^{\infty} f_k^{(+)} \left(|\rho - \rho_+|\right)^k,$$
$$G\big|_{\rho \to \rho_+} = \left(|\rho - \rho_+|\right)^{s_+} \sum_{k=0}^{\infty} g_k^{(+)} \left(|\rho - \rho_+|\right)^k. \tag{22}$$

The indicial equation for system (15) leads to the solution of

$$s_+ = -\frac{1}{2} \pm i \frac{\rho_+^2}{\rho_+ - \rho_-} \left( \varepsilon - \frac{\alpha_{em}}{\rho_+} \right). \tag{23}$$

At $\rho \to \rho_-$, let us present $F(\rho), G(\rho)$ as

$$F\big|_{\rho \to \rho_-} = \left(|\rho_- - \rho|\right)^{s_-} \sum_{k=0}^{\infty} f_k^{(-)} \left(|\rho_- - \rho|\right)^k,$$
$$G\big|_{\rho \to \rho_-} = \left(|\rho_- - \rho|\right)^{s_-} \sum_{k=0}^{\infty} g_k^{(-)} \left(|\rho_- - \rho|\right)^k. \tag{24}$$

In this case, the indicial equation for system (15) leads to the solution of

$$s_- = -\frac{1}{2} \pm i \frac{\rho_-^2}{\rho_+ - \rho_-} \left( \varepsilon - \frac{\alpha_{em}}{\rho_-} \right). \quad (25)$$

It is seen from (23), (25) that the oscillating parts of solutions (22), (24) disappear at fermion energies of $\varepsilon = \alpha_{em}/\rho_\pm$.

Formally, the solutions of $\varepsilon = \alpha_{em}/\rho_\pm$ are the unique solutions of equations (15). However, these solutions are non-physical due to the logarithmic divergence of the normalization integral

$$N_D = \int (F^* F + G^* G) \rho^2 d\rho \quad (26)$$

near the event horizons.

As the result, as well as the authors in [2], we arrive at the conclusion of absence of solutions to the Dirac equation corresponding to stationary bound states of fermions in the presence of event horizons in the Reissner-Nordström metric.

### 2.4.2 Extreme RN field $\left( \rho_+ = \rho_- = \alpha, \; \alpha^2 = \alpha_Q^2 \right)$

For the domain outside the event horizon $\rho > \alpha$ in [3], three conditions are specified, at fulfillment of which existence of stationary bound states of fermions is possible:

$$\varepsilon = \alpha_{em}/\alpha, \quad (27)$$

$$\kappa^2 + \alpha^2 - \alpha_{em}^2 > 1/4, \quad (28)$$

$$n + \sqrt{\alpha^2 - \alpha_{em}^2} + \sqrt{\kappa^2 + \alpha^2 - \alpha_{em}^2} = 0. \quad (29)$$

In (29), $n$ is a positive integer.

Since $|\varepsilon| < 1$ and $|\alpha_{em}| < \alpha$, the condition for the extreme RN field is not fulfilled. For the domain under the event horizon $0 < \rho < \alpha$, as well as in previous section 2.4.1, there is no possibility to formulate the problem for obtaining stationary solutions of the Dirac equation.

### 2.4.3 Naked RN singularity $\left( \alpha^2 < \alpha_Q^2 \right)$

At $\rho \to \infty$, asymptotics (21) is valid. At $\rho \to 0$, as well as in sections 2.4.1, 2.4.2, the problem of existence of two square integrable solutions of the Dirac equation persists.

So, absence of stationary solutions of the Dirac equation corresponding to bound fermion states in the RN field is shown in general. In presence of two event horizons, absence of stationary regular solutions is conditioned by square non-integrability of wave functions, which agrees with the conclusions of [2]. For the extreme RN field, absence of stationary bound states of fermions outside the event horizon is associated with non-fulfillment of condition (29) [3]. For the extreme RN field, there is no possibility in the domain under the event horizon to formulate the problem of determining stationary solutions to the Dirac equation due to presence of two square integrable solutions for radial wave functions [4]. The similar problem exists in case of naked RN singularity.

The situation qualitatively changes when using second-order equations with a spinor wave function for fermions.

### 3. Second-order equation for fermions in the Reissner-Nordström field

Obtaining the second-order equation involves three stages:
1. obtaining the self-conjugate Hamiltonian or the self-conjugate Dirac equation,
2. transition from bispinor to spinor wave functions in the second-order equation,
3. non-unitary similarity transformation to ensure self-conjugacy of the second-order equation with spinor wave functions.

The procedure for obtaining the second-order equation for fermions in the Reissner-Nordström field is presented in Appendix A. The equations for radial functions $\psi_F(\rho), \psi_G(\rho)$ have the form of the Schrödinger equation with effective potentials $U_{eff}^F(\rho), U_{eff}^G(\rho)$, non-linearly depending on energy $\varepsilon$

$$\frac{d^2\psi_F(\rho)}{d\rho^2} + 2\left(E_{Schr} - U_{eff}^F(\rho)\right)\psi_F(\rho) = 0, \tag{30}$$

$$\frac{d^2\psi_G(\rho)}{d\rho^2} + 2\left(E_{Schr} - U_{eff}^G(\rho)\right)\psi_G(\rho) = 0. \tag{31}$$

In (30), (31)

$$E_{Schr} = \frac{1}{2}\left(\varepsilon^2 - 1\right), \tag{32}$$

$$\psi_F(\rho) = g_F(\rho)F(\rho), \tag{33}$$

$$\psi_G(\rho) = g_G(\rho)G(\rho). \tag{34}$$

The explicit form of $U_{eff}(\rho)$ and $g(\rho)$ is presented in Appendixes A, B.

Equations (30), (31) transform into each other at $\varepsilon \to -\varepsilon, \kappa \to -\kappa, \alpha_{em} \to -\alpha_{em}$. It follows therefrom that equations (30), (31) describe the motion of particles and antiparticles. In this paper, equation (30) for function $\psi_F(\rho)$ with the effective potential $U_{eff}^F$ is used for particles. The non-relativistic limit of the Dirac equation with lower spinor, disappearing at zero momentum of particle $(\mathbf{p}=0)$ and proportional to $G(\rho)$, can serve as the basis for it. Similarly, the lower spinor with function $G(\rho)$ disappears for the particle in case of Foldy-Wouthuysen transformation with any value $\mathbf{p}$ [19]. Conversely, the upper spinor of the Dirac bispinor wave function, proportional to $F(\rho)$, disappears for the antiparticle within the non-relativistic limit $\mathbf{p} = 0$ and in case of Foldy-Wouthuysen transformation with any value $\mathbf{p}$.

### 3.1. Stationary solutions of $\varepsilon = \varepsilon_{RN}$

Here and below, $\varepsilon_{RN}$ are stationary solutions of second-order equation (30) $\varepsilon_{RN} = \alpha_{em}/\rho_+$, $\alpha_{em}/\rho_-$ in presence of two event horizons (see (23), (25)) and $\varepsilon_{RN} = \alpha_{em}/\alpha$ in case of the extreme RN field (see (27)).

#### 3.1.1 Singularities of effective potentials at $\varepsilon = \varepsilon_{RN}$

At $\rho \to \infty$,

$$U_{eff}^F(\varepsilon_{RN}) \to \varepsilon_{RN} \frac{\alpha_{em}}{\rho} + \frac{\alpha}{\rho}(1 - 2\varepsilon_{RN}^2) + O\left(\frac{1}{\rho^2}\right). \tag{35}$$

In presence of two event horizons $\rho_+, \rho_-$,

$$U_{eff}^F\left(\varepsilon = \frac{\alpha_{em}}{\rho_+}\right)\bigg|_{\rho \to \rho_+} = -\frac{3}{32}\frac{1}{(\rho - \rho_+)^2} + O\left(\frac{1}{|\rho - \rho_+|^{3/2}}\right), \tag{36}$$

$$U_{eff}^F\left(\varepsilon = \frac{\alpha_{em}}{\rho_-}\right)\bigg|_{\rho \to \rho_-} = -\frac{3}{32}\frac{1}{(\rho - \rho_-)^2} + O\left(\frac{1}{|\rho - \rho_-|^{3/2}}\right). \tag{37}$$

For the extreme RN field $(\rho_+ = \rho_- = \alpha, \alpha^2 = \alpha_Q^2)$,

$$U_{eff}^F\left(\varepsilon \neq \frac{\alpha_{em}}{\alpha}\right)\bigg|_{\rho \to \alpha} = -\frac{\alpha^4}{2(\rho - \alpha)^4}\left(\varepsilon - \frac{\alpha_{em}}{\alpha}\right)^2 + O\left(\frac{1}{|\rho - \alpha|^3}\right), \tag{38}$$

$$U_{eff}^F\left(\varepsilon = \frac{\alpha_{em}}{\alpha}\right)\bigg|_{\rho \to \alpha} = -\frac{1}{2(\rho - \alpha)^2}\left[\frac{1}{4} - (\kappa^2 + \alpha^2 - \alpha_{em}^2)\right] + O\left(\frac{1}{|\rho - \alpha|}\right). \tag{39}$$

Asymptotics (36), (37) represent inverse square potential wells with coefficient $K = 3/32 < 1/8$, which testifies to the possibility of existence of stationary bound states of quantum mechanical half-spin particles (see, for instance, [20]).

For the extreme RN field, the condition of existence of a potential well and the condition of existence of stationary bound states of Dirac particles $(K < 1/8)$ therein can be written from asymptotics (39) as

$$0 < \kappa^2 + \alpha^2 - \alpha_{em}^2 < 1/4. \tag{40}$$

Since for bound states $-1 < \varepsilon = \alpha_{em}/\alpha < 1$ and the constant of separation is $|\kappa| \geq 1$ (see (13)), condition (40) is not met at any admissible values of $\kappa, \alpha, \alpha_{em}$. Even at this stage, we can say that stationary bound states of fermions with energies $|\varepsilon| < 1$ do not exist in the extreme RN field.

In case of $\rho \to 0$ for any values $\varepsilon$,

$$U_{eff}^F \big|_{\rho \to 0} = \frac{3}{8\rho^2} + \mathrm{O}\left(\frac{1}{\rho}\right). \tag{41}$$

The asymptotics (41) represents an infinitely large repulsive barrier.

### 3.1.2 Square integrability of radial wave functions. Asymptotics of function $\psi_F(\rho, \varepsilon)$

**3.1.2.1** At $\rho \to \infty$, taking into account also $g_F\big|_{\rho \to \infty} = \rho$ (see Appendixes A, B)

$$\psi_F\big|_{\rho \to \infty} = \rho F\big|_{\rho \to \infty}. \tag{42}$$

For the finite motion of half-spin particles, taking into account (21),

$$\psi_F\big|_{\rho \to \infty} = C_1 \varphi_1(\rho) \rho e^{-\sqrt{1-\varepsilon^2}\rho}. \tag{43}$$

**3.1.2.2** In presence of two event horizons, let us present the function of $\psi_F(\rho, \varepsilon_{RN})$ in the following form:

at $\rho \to \rho_+$

$$\psi_F\left(\varepsilon = \frac{\alpha_{em}}{\rho_+}\right)\bigg|_{\rho \to \rho_+} = |\rho - \rho_+|^s \sum_{k=0}^{\infty} \chi_k^{(+)} |\rho - \rho_+|^k ; \tag{44}$$

at $\rho \to \rho_-$

$$\psi_F\left(\varepsilon = \frac{\alpha_{em}}{\rho_-}\right)\bigg|_{\rho \to \rho_-} = |\rho_- - \rho|^s \sum_{k=0}^{\infty} \chi_k^{(-)} |\rho_- - \rho|^k . \tag{45}$$

From (30), taking into account (44), (45), (36), (37) the indicial equation of

$$s(s-1) + 3/16 = 0 \tag{46}$$

with the solutions of $s_1 = 3/4$, $s_2 = 1/4$ follows.

Both the solutions lead to regular square integrable solutions for the wave function $\psi_F(\rho, \varepsilon_{RN})$. For the unambiguous choice of the solution, let us turn to asymptotics (22), (24) for the radial function of the Dirac equation $F(\rho)$ and to transformation (33) with $g_F\big|_{\rho_\pm} \sim |\rho - \rho_\pm|^{3/4}$ (see Appendixes A, B). We obtain

$$\psi_F\left(\varepsilon = \frac{\alpha_{em}}{\rho_+}\right)\bigg|_{\rho \to \rho_+} = C_3 |\rho - \rho_+|^{1/4}, \quad (47)$$

$$\psi_F\left(\varepsilon = \frac{\alpha_{em}}{\rho_-}\right)\bigg|_{\rho \to \rho_-} = C_4 |\rho_- - \rho|^{1/4}. \quad (48)$$

The asymptotics (47), (48) correspond to the solution of indicial equation (46) of $s_2 = 1/4$. Below, we will use the solutions of equation (30) with asymptotics (47), (48) as eigenfunctions of stationary bound states of fermions with eigenvalues $\varepsilon_{RN} = \alpha_{em}/\rho_\pm$. These solutions are square integrable in the neighborhood of event horizons. Note that wave functions (47), (48) on the event horizons $\rho_+, \rho_-$ are zero.

**3.1.2.3** At $\rho \to 0$, for any value of $\varepsilon$, let us present function $\psi_F(\rho,\varepsilon)$ as

$$\psi_F(\varepsilon)\big|_{\rho \to 0} = \rho^s \sum_{k=0} \chi_k^{(0)} \rho^k. \quad (49)$$

From (30), taking into account (49), (41) the indicial equation has the form of

$$s(s-1) - 3/4 = 0. \quad (50)$$

with the solutions of $s_1 = 3/2$, $s_2 = -1/2$.

We reject the solution with $s_2$, since it leads to absence of square integrability of $\psi_F(\rho,\varepsilon)$ in the neighborhood of $\rho = 0$.

As the result, we have

$$\psi_F(\varepsilon)\big|_{\rho \to 0} = C_5 \rho^{3/2}. \quad (51)$$

So, we have found out the unambiguous asymptotic behavior of the wave functions of second-order equation (30) in presence of two event horizons $\rho_+, \rho_-$ ($\rho \in (0,\rho_-]$, $\rho \in [\rho_+,\infty)$) and for naked singularity $\rho \in (0,\infty)$. Likewise, we can find out the asymptotics of wave functions in case of the extreme RN field ($\rho_+ = \rho_- = \alpha$; $\alpha^2 = \alpha_Q^2$). However, we do not need it for our purposes, since we have shown above the absence of stationary bound states of fermions with $|\varepsilon| < 1$ for this case on both sides of the single event horizon.

### 3.1.3 Domain of radial wave functions $\psi_F(\rho,\varepsilon)$

For the naked RN singularity, the following domain is the domain of radial wave functions

$$\rho \in (0,\infty). \quad (52)$$

In presence of two event horizons, we can split domain (52) into the three domains of:

$$\rho \in [\rho_+, \infty), \tag{53}$$

$$\rho \in (\rho_-, \rho_+), \tag{54}$$

$$\rho \in (0, \rho_-]. \tag{55}$$

Domains (53), (54) and domains (54), (55) are separated from each other by infinitely deep potential wells of $\sim -\frac{3}{32}\frac{1}{(\rho-\rho_\pm)^2}$ (see (36), (37)). The wave functions on the event horizon are zero (see (47), (48)). In domains (53), (55), the effective potential of $U_{eff}^F(\rho)$ and radial wave function $\psi_F(\rho)$ are real. In domain (54), between event horizons, these values become complex. Below, we will examine domains (53), (55). For these domains, regular real square integrable radial wave functions of $\psi_F(\varepsilon_{RN}, \rho)$, vanishing at $\rho = \rho_+$ or $\rho = \rho_-$, correspond to the solutions of the second-order equations $\varepsilon = \alpha_{em}/\rho_+$, $\varepsilon = \alpha_{em}/\rho_-$.

We will demonstrate this in section 4 by numerical solutions of second-order equation (30).

### 3.2 Impenetrable potential barriers

Effective potential (A.22), at some $\rho = \rho_{cl}^\pm$, can have singularities of the following form

$$U_{eff}^F\Big|_{\rho \to \rho_{cl}^\pm} \sim \frac{1}{\left(\varepsilon - \frac{\alpha_{em}}{\rho} + \sqrt{1 - \frac{2\alpha}{\rho} + \frac{\alpha_Q^2}{\rho^2}}\right)^2}. \tag{56}$$

These singularities can be contained in the second summand (A.22), equal to $\frac{3}{8}\frac{1}{B^2}\left(\frac{dB}{d\rho}\right)^2$ (see Appendix A).

The radii $\rho_{cl}^\pm$, at which the expression in the denominator (56) is zero, are determined by equalities of

$$\rho_{cl}^\pm = \frac{\alpha - \alpha_{em}\varepsilon \pm \sqrt{(\alpha - \alpha_{em}\varepsilon)^2 - (\alpha_Q^2 - \alpha_{em}^2)(1-\varepsilon^2)}}{1-\varepsilon^2}. \tag{57}$$

Asymptotic (56), taking into account (57), can be presented as

$$U_{eff}^F\Big|_{\rho \to \rho_{cl}^\pm} = \frac{3}{8(\rho - \rho_{cl}^\pm)^2} + O\left(\frac{1}{|\rho - \rho_{cl}^\pm|}\right). \tag{58}$$

Such potential barriers are known to be impenetrable [21][2]

The potential well necessary for emergence of bound states with $-1 < \varepsilon_n < 1$ is absent at existence of at least one singularity (58) in effective potential $U_{eff}^F(\rho)$. Fig. 1 presents characteristic curves of $U_{eff}^F(\rho)$ with one or two singularities.

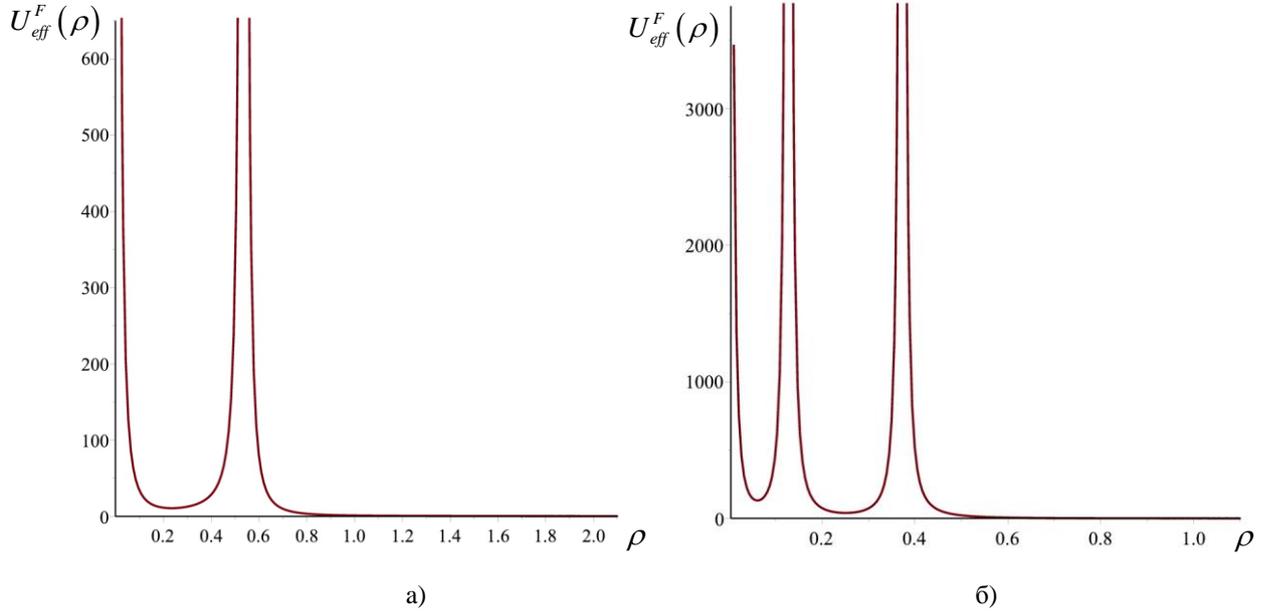

a)            б)

Fig. 1. The curves of $U_{eff}^F(\rho)$ in the RN field of naked singularity at

a) $\alpha = 0.25;\ \alpha_Q = 0.5;\ \alpha_{em} = 1;\ \kappa = -1;\ \varepsilon = 0.9;\ \rho_{cl}^+ \simeq 0.535$;

b) $\alpha = 0,25;\ \alpha_Q = 0,5;\ \alpha_{em} = 0,45;\ \kappa = -1;\ \varepsilon = 0;\ \rho_{cl}^- = 0,125;\ \rho_{cl}^+ = 0,375$.

Then, let us consider the conditions of singularity emergence (58) in presence of two event horizons and in case of naked RN singularity.

**3.2.1 Availability of two event horizons** $\rho_+, \rho_-\ (\alpha^2 > \alpha_Q^2)$; **domains of wave functions:** $\rho \in [\rho_+, \infty),\ \rho \in (0, \rho_-]$

Firstly, the expression

$$\varepsilon - \frac{\alpha_{em}}{\rho} + \sqrt{\frac{(\rho - \rho_+)(\rho - \rho_-)}{\rho^2}} \tag{59}$$

can be equal to zero at $\varepsilon = 0$, $\alpha_{em} = 0$ and $\rho = \rho_\pm$. In this case, no new singularity $U_{eff}^F$ appears. As before,

$$U_{eff}^F\Big|_{\rho \to \rho_\pm} = -\frac{3}{32}\frac{1}{(\rho - \rho_\pm)^2} \quad \text{(see (36), (37))}.$$

---

[2] Note that the authors [21] used the Schrödinger-type equation (30) without factor 2. In our case, barrier $K/(\rho - \rho_{cl}^\pm)^2$ is impenetrable if $K \geq 3/8$.

Secondly, let us consider the possibility of vanishing for expression (59) at $\varepsilon = \alpha_{em}/\rho_{\pm}$.

$$\sqrt{\frac{(\rho_{cl}-\rho_{+})(\rho_{cl}-\rho_{-})}{\rho_{cl}^2}} = \frac{\alpha_{em}}{\rho_{cl}} - \varepsilon = \alpha_{em}\left(\frac{\rho_{\pm}-\rho_{cl}}{\rho_{cl}\rho_{\pm}}\right). \tag{60}$$

For domain $\rho \in [\rho_{+},\infty)$, we look for existence of $\rho_{cl} > \rho_{+}$. It follows from the positivity of the left part of equality (60) that it is possible only at $\alpha_{em} < 0$ and $-1 < \varepsilon < 0$. For the domain of $\rho \in (0, \rho_{-})$, a possible value of $\rho_{cl}$ should be lower than $\rho_{-}$. It is possible from equality (60) only at $\alpha_{em} > 0$ and $0 < \varepsilon < 1$. It is possible to show from equality (60) that inequalities of $\rho_{cl} > \rho_{+}$ and $\rho_{cl} < \rho_{-}$ are not fulfilled for both the domains of the wave functions. So, to solve $\varepsilon = \alpha_{em}/\rho_{\pm}$ for bound states of $-1 < \varepsilon < 1$, expression (56) is non-singular and there are no impenetrable barriers of form (58).

### 3.2.2 Naked RN singularity $(\alpha_Q^2 > \alpha^2)$; domain of wave functions: $\rho \in (0, \infty)$

As above, we consider the conditions of existence $\rho_{cl}$ for bound states with energies within the range of $-1 < \varepsilon < 1$.

From the positivity of $\sqrt{1 - \frac{2\alpha}{\rho} + \frac{\alpha_Q^2}{\rho^2}}$, the equality of the denominator to zero in (56) is possible only at

$$(\alpha_{em}/\rho) - \varepsilon > 0. \tag{61}$$

Inequality (61) for $\rho = \rho_{cl}^{\pm}$ can be presented in the following view.

In the case $\alpha_{em} > 0$ (like charges of the field source of naked RN singularity and the fermion):

$$(\varepsilon/\alpha_{em})\rho_{cl}^{\pm} < 1 \text{ at } 0 < \varepsilon < 1, \tag{62}$$

$$-(|\varepsilon|/\alpha_{em})\rho_{cl}^{\pm} < 1 \text{ at } -1 < \varepsilon < 0. \tag{63}$$

Inequality (63) is fulfilled at any value of $\alpha_{em} > 0$ and $|\varepsilon| < 1$.

In the case $\alpha_{em} < 0$ (opposite charges of the field source of the naked RN singularity and the fermion):

$$\frac{|\varepsilon|}{|\alpha_{em}|}\rho_{cl}^{\pm} > 1 \text{ at } -1 < \varepsilon < 0. \tag{64}$$

For the range of $0 < \varepsilon < 1$ and $\alpha_{em} < 0$, inequality (61) is not fulfilled and there are no singularities like (58).

The positivity condition of the radical expression of $N \geq 0$ in (57) can be presented as

$$N = \alpha_Q^2 \varepsilon^2 - 2\alpha\alpha_{em}\varepsilon + \alpha^2 - \alpha_Q^2 + \alpha_{em}^2 \geq 0. \quad (65)$$

The discriminant of the trinomial (65) is $D = 4\left(\alpha_{em}^2 - \alpha_Q^2\right)\left(\alpha_Q^2 - \alpha^2\right)$.

The condition (65) is fulfilled if $D \geq 0$, i.e. $\alpha_{em}^2 \geq \alpha_Q^2$; condition (65) is partially fulfilled, if $D < 0$ and $\alpha_{em}^2 < \alpha_Q^2$. In this case, the roots of trinomial (65) are

$$\varepsilon_{\pm} = \frac{\alpha\alpha_{em} \pm \sqrt{\left(\alpha_Q^2 - \alpha_{em}^2\right)\left(\alpha_Q^2 - \alpha^2\right)}}{\alpha_Q^2}. \quad (66)$$

Condition (65) is met at $D < 0$, if $\varepsilon \notin (\varepsilon_-, \varepsilon_+)$.

As the result, the analysis of singularity existence of type (58) is reduced to fulfillment of the three inequalities:

$$\left(\alpha_{em}/\rho_{cl}^{\pm}\right) - \varepsilon > 0; \quad \rho_{cl}^{\pm} \geq 0; \quad N \geq 0. \quad (67)$$

The results of the analysis are presented in Table 1.

Table1. Conditions of existence of impenetrable barriers and bound states of fermions in the field of naked RN singularity

| Ranges of $\alpha_{em}$ and $\varepsilon$ | | Conditions | $\rho_{cl}^+$ | $\rho_{cl}^-$ |
|---|---|---|---|---|
| $\alpha_{em} > 0$, $0 < \varepsilon < 1$ | $\varepsilon < \min(\alpha_{em}/\alpha,\ \alpha/\alpha_{em})$ | $\alpha_{em}^2 \geq \alpha_Q^2$ | ∃ | ∄ |
| | | $\alpha_{em}^2 < \alpha_Q^2$ at $\varepsilon \notin [\varepsilon_-, \varepsilon_+]$ | ∃ | ∃ |
| | | $\alpha_{em}^2 < \alpha_Q^2$ at $\varepsilon \in [\varepsilon_-, \varepsilon_+]$ | ∄ | ∄ |
| | $\varepsilon > \alpha/\alpha_{em}$ | $\alpha_{em}^2 \geq \alpha_Q^2$ | ∃ | ∄ |
| | | $\alpha_{em}^2 < \alpha_Q^2$ | ∄ | ∄ |
| | $\varepsilon > \alpha_{em}/\alpha$ | without conditions | ∄ | ∄ |
| $\alpha_{em} > 0$, $\varepsilon = 0$ | | $\alpha_{em}^2 \geq \alpha_Q^2 - \alpha^2$ | ∃ | |
| | | $\alpha_Q^2 > \alpha_{em}^2 > \alpha_Q^2 - \alpha^2$ | ∃ | ∃ |
| | | $\alpha_{em}^2 < \alpha_Q^2 - \alpha^2$ | ∄ | ∄ |
| $\alpha_{em} > 0$, $-1 < \varepsilon < 0$ | | $\alpha_{em}^2 \geq \alpha_Q^2$ | ∃ | ∄ |
| | | $\alpha_{em}^2 < \alpha_Q^2$ at $\varepsilon \notin [\varepsilon_-, \varepsilon_+]$ | ∃ | ∃ |
| | | $\alpha_{em}^2 < \alpha_Q^2$ at $\varepsilon \in [\varepsilon_-, \varepsilon_+]$ | ∄ | ∄ |
| $\alpha_{em} < 0$, $0 < \varepsilon < 1$ | | without conditions | ∄ | ∄ |
| $\alpha_{em} < 0$, $-1 < \varepsilon < 0$ | $|\varepsilon| < |\alpha_{em}|/\alpha$ | without conditions | ∄ | ∄ |
| | $|\varepsilon| > |\alpha_{em}|/\alpha$ | $\alpha_{em}^2 \geq \alpha_Q^2$ | ∃ | ∄ |
| | | $\alpha_{em}^2 < \alpha_Q^2$ at $\varepsilon \notin [\varepsilon_-, \varepsilon_+]$ | ∃ | ∃ |
| | | $\alpha_{em}^2 < \alpha_Q^2$ at $\varepsilon \in [\varepsilon_-, \varepsilon_+]$ | ∄ | ∄ |

So, it follows from the analysis that the existence of stationary bound states of fermions in the field of naked RN singularity is possible at definite values of physical parameters. Let us note that at $\alpha_{em} < 0$ (opposite charges) and $|\alpha_{em}| < \alpha$, the parameters of the RN metric cannot provide for existence of tightly coupled stationary states of fermions with $-1 < \varepsilon < 0$.

## 4. Prüfer transformation and asymptotic of functions

For numerical solution of the second-order equation, let us apply the Prüfer transformation [7] - [10] to equation (30) with the effective potential of $U_{eff}^F(\rho)$

$$\psi_F(\rho) = P(\rho)\sin\Phi(\rho),$$
$$\frac{d\psi_F(\rho)}{d\rho} = P(\rho)\cos\Phi(\rho). \tag{68}$$

Then,

$$\psi_F(\rho) \bigg/ \frac{d\psi_F(\rho)}{d\rho} = \text{tg}\,\Phi(\rho) \tag{69}$$

And we can write equation (30) as the system of first-order non-linear differential equations

$$\frac{d\Phi}{d\rho} = \cos^2\Phi + 2(E_{Schr} - U_{eff}^F)\sin^2\Phi, \tag{70}$$

$$\frac{d\ln P}{d\rho} = \left(1 - 2(E_{Schr} - U_{eff}^F)\right)\sin\Phi\cos\Phi. \tag{71}$$

Note, that the equation (71) should be solved upon determination of eigenvalues $\varepsilon_n$ and eigenfunctions $\Phi_n(\rho)$ from equation (70).

### 4.1 Asymptotics of functions $\Phi(\rho)$, $P(\rho)$

**4.1.1** For bound states at $\rho \to \infty$, taking into account (43), (69), we obtain that

$$\text{tg}\,\Phi\big|_{\rho\to\infty} = -\frac{1}{\sqrt{1-\varepsilon^2}},$$
$$\Phi\big|_{\rho\to\infty} = -\text{arctg}\frac{1}{\sqrt{1-\varepsilon^2}} + k\pi. \tag{72}$$

For exponentially increasing solutions in asymptotics (21), $C_2 \neq 0$ and, taking into account (42), (69), we have

$$\text{tg}\,\Phi\big|_{\rho\to\infty} = \frac{1}{\sqrt{1-\varepsilon^2}},$$
$$\Phi\big|_{\rho\to\infty} = \text{arctg}\frac{1}{\sqrt{1-\varepsilon^2}} + k\pi. \tag{73}$$

In (72), (73) $k = 0, \pm 1, \pm 2, \ldots$

**4.1.2** Let us first examine the case of $\rho \to \rho_+$ in presence of two event horizons.

Let
$$\Phi\big|_{\rho \to \rho_+} = k\pi + A|\rho - \rho_+|. \tag{74}$$

Then, $\sin\Phi\big|_{\rho \to \rho_+} \simeq \pm A|\rho - \rho_+|$; $\cos\Phi\big|_{\rho \to \rho_+} \simeq \pm 1$.

It follows from the consistency (74) with equation (70), taking into account the leading singularity, that $U_{eff}^F\left(\varepsilon = \dfrac{\alpha_{em}}{\rho_+}\right)\bigg|_{\rho \to \rho_+} = -\dfrac{3}{32}\dfrac{1}{(\rho - \rho_+)^2}$ (see (36))

$$1 + \frac{3}{16}A^2 = A \tag{75}$$

with the solutions of $A_1 = 4$; $A_2 = 4/3$. Then, let us integrate equation (71) at $\rho \to \rho_+$, taking into account the leading singularity of effective potential (36). As the result,

$$P\big|_{\rho \to \rho_+} = C_6 \begin{cases} |\rho - \rho_+|^{-3/4}, A_1 = 4, \\ |\rho - \rho_+|^{-1/4}, A_2 = 4/3, \end{cases} \tag{76}$$

$$\psi_F\left(\varepsilon = \frac{\alpha_{em}}{\rho_+}\right)\bigg|_{\rho \to \rho_+} = C_6 \begin{cases} 4|\rho - \rho_+|^{1/4}, A_1 = 4, \\ \dfrac{4}{3}|\rho - \rho_+|^{3/4}, A_2 = 4/3. \end{cases} \tag{77}$$

The comparison with (47) shows that solutions (74), (76), (77) with the solution of equation (75) $A_1 = 4$ and $C_3 = 4C_6$ are acceptable for our consideration.

The similar examination at $\rho \to \rho_-$ leads to the following physically acceptable asymptotics for $\Phi(\rho)$ and $P(\rho)$

$$\Phi\big|_{\rho \to \rho_-} = -4|\rho_- - \rho| + k\pi, \tag{78}$$

$$P\big|_{\rho \to \rho_-} = -\frac{C_4}{4}|\rho_- - \rho|^{-3/4}, \tag{79}$$

$$\psi_F\left(\varepsilon = \frac{\alpha_{em}}{\rho_-}\right)\bigg|_{\rho \to \rho_-} = C_4|\rho_- - \rho|^{1/4}. \tag{80}$$

**4.1.3** Let us consider the asymptotics of $\Phi(\rho), P(\rho)$ at $\rho \to 0$.

Let
$$\Phi\big|_{\rho \to 0} = k\pi + B\rho. \tag{81}$$

Then, from the equation (70), taking into account the leading singularity of (41) $U_{eff}^F\big|_{\rho \to 0} = \dfrac{3}{8}\dfrac{1}{\rho^2}$, we obtain the following equation

$$1 - \frac{3}{4}B^2 = B \tag{82}$$

with the solutions of $B_1 = 2/3;\ B_2 = -2$.

Upon integration (71) at $\rho \to 0$, taking into the leading singularity (41), we obtain that

$$P\big|_{\rho \to 0} = C_7 \begin{cases} \rho^{1/2},\ B_1 = 2/3, \\ \rho^{-3/2},\ B_2 = -2, \end{cases} \tag{83}$$

$$\psi_F\big|_{\rho \to 0} = C_7 \begin{cases} \dfrac{2}{3}\rho^{3/2},\ B_1 = 2/3, \\ -2\rho^{-1/2},\ B_2 = -2. \end{cases} \tag{84}$$

The comparison with (51) shows that solutions (81), (83), (84) with the solution of equation (82) $B_1 = 2/3$ and with $C_5 = (2/3)C_7$ are physically acceptable.

**5. Numerical method of solving equations for phase functions. General properties of phase functions**

In this paper, we use the following numerical method for solving equation (70).

For the permitted set of values of $-1 < \varepsilon < 1$, we numerically solve the Cauchy problem with the specified initial condition. To solve the Cauchy problem, we use the fifth-order Runge-Kutta implicit method with step control (the Ehle scheme of the Radau II A three-stage method [22]).

Determining the spectrum of $\varepsilon_n$ and eigenfunctions of $\Phi_n(\rho)$ by the solution (70) and integrating equation (71), one can determine functions of $P_n(\rho)$ and wave functions $(\psi_F)_n(\rho)$, taking into account (68). Then we can determine the probability density of fermion detection in the state with $\varepsilon_n$ at the distance $\rho$ in the spherical layer $d\rho$

$$w(\rho) = P_n^2(\rho)\sin^2 \Phi_n(\rho) \tag{85}$$

and detection probability of bound fermions within the range of $[\rho_0, \rho]$

$$W(\rho) = \int_{\rho_0}^{\rho} P_n^2(\rho')\sin^2 \Phi_n(\rho')d\rho'. \tag{86}$$

In presence of two event horizons, the energy of bound states is determined by equalities of $\varepsilon = \alpha_{em}/\rho_\pm$. In this case, only eigenvalues of function $(\psi_F)_n(\rho)$ (68), probability density (85) and integral probabilities (86) are determined numerically.

The wave functions $(\psi_F)_n(\rho)$ should satisfy asymptotics (43), (77) with $A_1 = 4$, (80) with $A_1 = -4$, (84) with $B_1 = 2/3$.

When solving the equation (70), it is necessary to keep in mind the existence of three irregular singular points: $\rho = 0$, $\rho = \rho_+$, $\rho = \rho_-$. The numerical calculations started and ended in the $\delta$-neighborhood of irregular singular points $\rho = 0$, $\rho = \rho_+$, $\rho = \rho_-$ with $\delta = 10^{-8}$ with good convergence of the results. The choice of the maximal value of $\rho_{max}$ in the calculations with simulation of $\rho \to \infty$ was determined by fulfillment of conditions (72), (73) within the specified accuracy of $10^{-7}$.

Below, we will use the function of $\Phi(\varepsilon, \rho_{max}) = \Phi(\varepsilon)|_{\rho = \rho_{max}}$ for the case of naked RN singularity while determining the spectrum $\varepsilon_n$. Here, $\rho_{max}$ is the maximal distance in numerical calculations. As a rule, the value of $\rho_{max} = 10^7$ ensures good convergence of the results.

The numerical calculations have shown availability of the following essential properties of the function $\Phi(\varepsilon, \rho_{min})$ (similar properties of the function $\Phi$ for simpler potentials, not depending on $\varepsilon$, were rigorously proved in [8] - [10]).

1. Function $\Phi(\varepsilon, \rho_{max})$ is $\varepsilon$ monotonous.

2. In case of existence of bound states with $-1 < \varepsilon < 1$, the behavior of $\Phi(\varepsilon, \rho_{max})$ is stepwise. Upon achieving the eigenvalue of $\varepsilon_n$, the function of $\Phi(\varepsilon, \rho_{max})$ varies stepwise by $\pi$

$$\left[\Phi(\varepsilon_0 - \Delta\varepsilon, \rho_{max}) - \Phi(\varepsilon_n + \Delta\varepsilon, \rho_{max})\right]\Big|_{\Delta\varepsilon \to 0} = \pm n\pi. \quad (87)$$

3. In case of absence of bound states, the variations in the function of $\Phi(\varepsilon, \rho_{max})$ is less than the value $\pi$ within the entire range of $-1 < \varepsilon < 1$.

**5.1 Availability of two event horizons $\rho_+, \rho_-$; $\alpha^2 > \alpha_Q^2$. Domain of wave functions $\rho \in [\rho_+, \infty)$**

In this case, the $m_\varphi$-degenerate solution (36) exists: $\varepsilon = \alpha_{em}/\rho_+$. For the bound states, $-1 < \varepsilon < 1$, therefore $-\rho_+ < \alpha_{em} < \rho_+$. The solution of $\varepsilon = \alpha_{em}/\rho_+$ involves states with like and opposite charges of the RN field source and the fermion. The case of $\varepsilon = 0$, $\alpha_{em} = 0$ corresponds to an uncharged half-spin particle.

When determining wave functions with the known eigenfvalue of $\varepsilon = \alpha_{em}/\rho_+$, equation (70) is integrated from the right to the left (from $\rho = \rho_{max}$ with boundary conditions (72) to $\rho = \rho_+$ with asymptotics (74) and with solution of equation (75) $A_1 = 4$)).

Fig. 2 presents the family of integral curves of equation (70) near the event horizon $\rho_+$, being the irregular singular point of this equation. The separatrices (red curves), corresponding to asymptotics (74) - (77), are shown with $A_1 = 4$ and $A_2 = 4/3$. Two separatrices begin at $(\rho_+, 0)$. In the neighborhood of event horizon $\rho_+$, choice of the "physical" separatrix with $A_1 = 4$ is realized in the calculations while integrating (70) from the "right to the left".

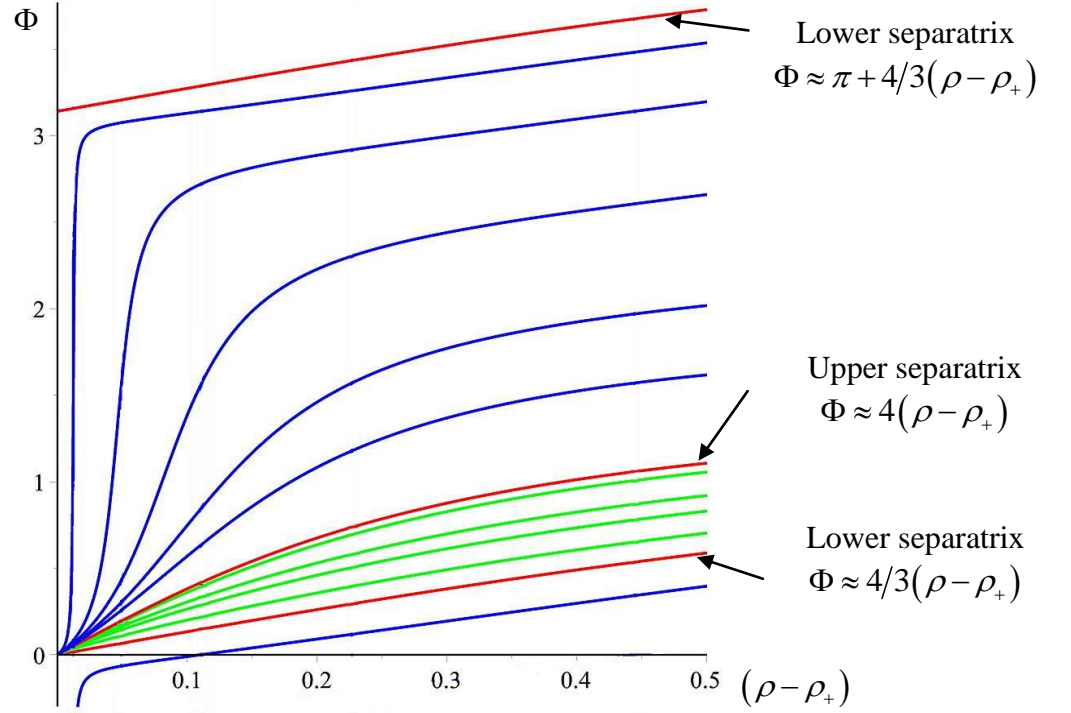

Fig. 2. The integral curves of $\Phi(\rho - \rho_+)$ in the neighborhood of $\rho_+$.

### 5.1.1 Determining the boundaries of physical acceptability of the solution $\varepsilon = \alpha_{em}/\rho_+$ and calculated results

Solution $\varepsilon = 0$ at $\alpha_{em} = 0$ is formally valid for any value of a gravitational coupling constant $\alpha$. However, in order to achieve such a strong coupling of $\varepsilon_b = 1$, the values of $\alpha \geq \alpha_{min}$ are needed, as well as in case of the Schwarzschild field [1]. For instance, the energy of a bound electron $\varepsilon_e = 0$ in the Coulomb field of the nucleus with the number of protons $Z$ in the $1S_{1/2}$ state is achieved at the value of the electromagnetic coupling constant of $\alpha_{fs} Z \simeq Z/137 \approx 1.06$ $(Z = 140)$. For the states of $2S_{1/2}$ and $3S_{1/2}$, the similar values are $\alpha_{fs} Z \simeq 1.42$ and 1.9 [23].

In [1], it is determined for the Schwarzschild field that
$$\alpha_{min} \sim 0,25. \qquad (88)$$

Hereafter, the symbol of "~" stands for $\alpha_{\min} \approx 0.25$, but always $\alpha_{\min} > 0.25$.

At $\alpha_Q = 0, \alpha_{em} = 0$, the RN field transforms into the Schwarzschild field. The maximal value of $|\alpha_Q|^{\max} = \alpha$ in presence of event horizons corresponds to the extreme RN field. Since there are no stationary bound states of fermions for the extreme RN field, in our denotations,

$$\alpha_{\min} \sim |\alpha_Q|^{\max}. \qquad (89)$$

For bound states of $-1 < \varepsilon < 1$ and for solutions of $\varepsilon = \alpha_{em}/\rho_{\pm}$, the following condition should be met

$$\rho_{\pm} \sim \alpha_{em}. \qquad (90)$$

In Table 2, for solution $\varepsilon = 0$, $\alpha_{\min} = 0.251$, $\alpha_Q/\alpha_{\min} = 0, 0.5, 0.9$ and for different values of $j, l$, the computed distances $\rho_m$ from the maxima of probability densities to the event horizons of $\rho_+$ are presented. Fig. 3 presents normalized probability densities (85) and integral probabilities (86) for $\kappa = -1$ $(j = 1/2, l = 0)$. Table 3 presents the values of $\rho_m$ versus $\alpha_{em}$ for $\alpha_{\min} = 0.251$.

Table 2. The values of $\rho_m$ for $\varepsilon = 0$, $\alpha_{\min} = 0.251$, $\alpha_Q/\alpha_{\min} = 0, 0.5, 0.9$ versus $\kappa(j,l)$.

|  | $\alpha_Q/\alpha_{\min}$ | 0 | 0.5 | 0.9 |
|---|---|---|---|---|
| $\rho_m$ | $\kappa = -1$ $(j=1/2, l=0)$ | 0.043 | 0.04 | 0.023 |
|  | $\kappa = +1$ $(j=1/2, l=1)$ | 0.023 | 0.021 | 0.012 |
| $\rho_+$ |  | 0.502 | 0.4684 | 0.3604 |

Table 3. The values of $\rho_m$ for $\alpha_{\min} = 0.251$, $\alpha_Q/\alpha_{\min} = 0.5$, $\rho_+ = 0.4684$ versus $\alpha_{em}$ and $\kappa(j,l)$.

| $\kappa$ | $-1$ | | | | | $+1$ | | | | |
|---|---|---|---|---|---|---|---|---|---|---|
| $(j,l)$ | $(1/2, 0)$ | | | | | $(1/2, 1)$ | | | | |
| $\alpha_{em}$ | $+0.4679$ | $+0.2342$ | $0$ | $-0.2342$ | $-0.4679$ | $+0.4679$ | $+0.2342$ | $0$ | $-0.2342$ | $-0.4679$ |
| $\varepsilon$ | $+0.999$ | $+0.5$ | $0$ | $-0.5$ | $-0.999$ | $+0.999$ | $+0.5$ | $0$ | $-0.5$ | $-0.999$ |
| $\rho_m$ | 0.059 | 0.045 | 0,04 | 0.033 | 0.02 | 0.016 | 0.018 | 0.021 | 0.026 | 0.037 |
| $\kappa$ | $-2$ | | | | | $+2$ | | | | |
| $(j,l)$ | $(3/2, 1)$ | | | | | $(3/2, 2)$ | | | | |
| $\alpha_{em}$ | $+0.4679$ | $+0.2342$ | $0$ | $-0.2342$ | $-0.4679$ | $+0.4679$ | $+0.2342$ | $0$ | $-0.2342$ | $-0.4679$ |
| $\varepsilon$ | $+0.999$ | $+0.5$ | $0$ | $-0.5$ | $-0.999$ | $+0.999$ | $+0.5$ | $0$ | $-0.5$ | $-0.999$ |
| $\rho_m$ | 0.0097 | 0.0089 | 0.008 | 0.0069 | 0.0057 | 0.0052 | 0.0057 | 0.0063 | 0.0072 | 0.0084 |

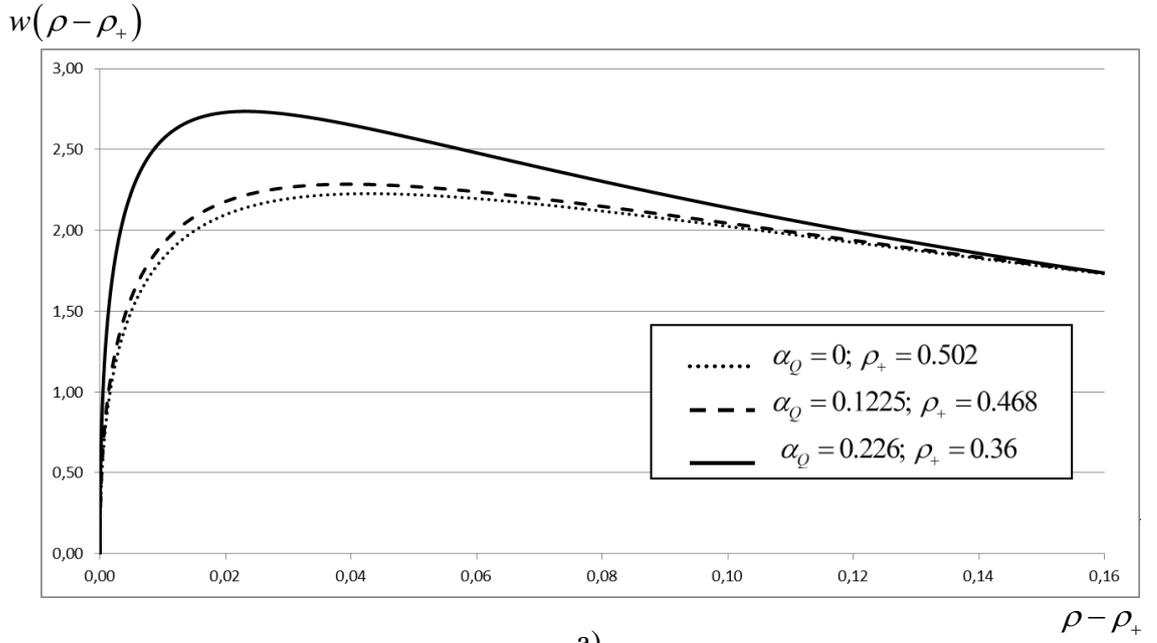

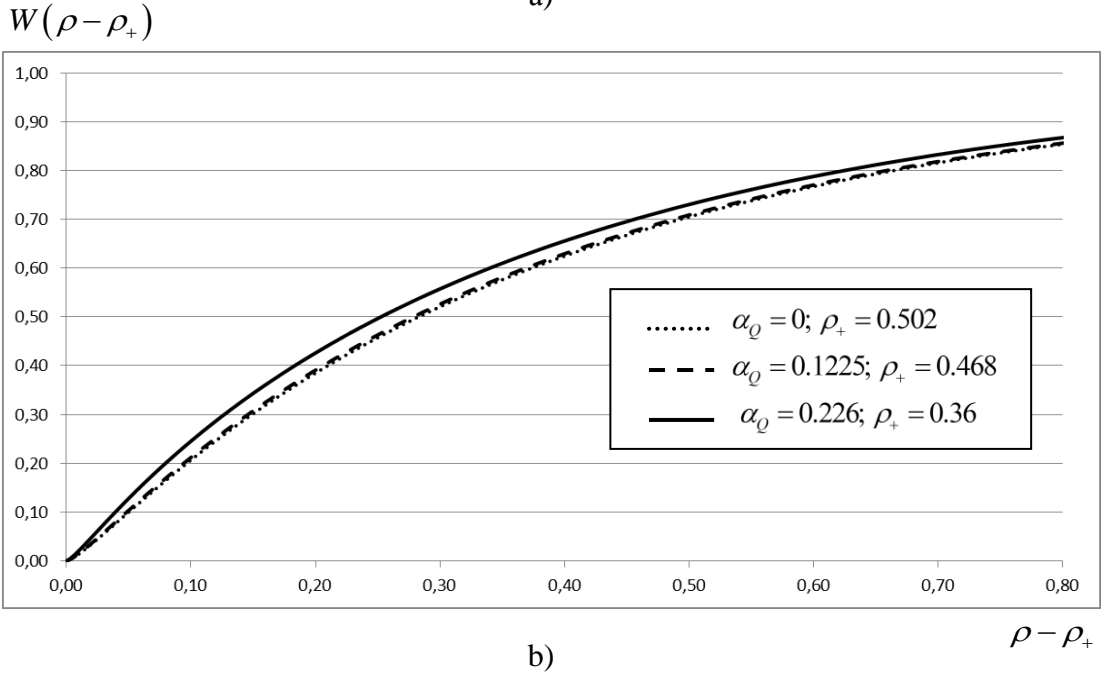

Fig. 3. The curves: a) $w(\rho-\rho_+)$, b) $W(\rho-\rho_+)$ for bound states with $\varepsilon=0$, $\alpha_{\min}=0.251$, $\kappa=-1$ $(j=1/2, l=0)$ for values of $\alpha_Q/\alpha_{\min}=0; 0.5; 0.9$.

The calculations show that the localization of particles near the event horizons of $\rho_+$, as distinct from the electronic structure of atoms in the periodic table, increases with increase in $j,l$. The data of Tables 2, 3, and Fig.3 show that localization of particles near $\rho_+$ increases with increase in $|\alpha_Q|$. For the states of $j=l+1/2$ $(\kappa<0)$, the maximal values of $\rho_m$ at $\alpha_{em}>0$ are achieved at $\rho_+ \sim \alpha_{em}^{\max}$. For opposite charges of the RN field source and the fermion $(\alpha_{em}<0)$, the minimal values of $\rho_m$ are achieved at the minimal admissible value of $\alpha_{em}^{\min} \sim -\rho_+$.

Conversely, for the states of $j = l - 1/2$ $(\kappa > 0)$, the maximal values of $\rho_m$ are achieved at $\alpha_{em}^{max} \sim -\rho_+$, and the minimal values are achieved at $\alpha_{em}^{max} \sim \rho_+$.

The solution of $\varepsilon = \alpha_{em}/\rho_+$ can be implemented only at definite ratios between masses and charges of the RN field source and the fermion.

1. We assume that $M \geq M_P$

2. For existence of event horizons, $\alpha_Q = k\alpha$, where $0 < k < 1$

3. For simplicity, we consider the ratio of masses of $m/M \leq 10^{-2}$. In this case, motion of fermions can be described relative to the RN field source at rest. For large masses of fermions, the motion of the RN field source should be taken into account, that is a complicate and cumbersome problem in the curved space-time (see, for instance [24]).

4. Existence of the limiting value of $\rho_+ \sim |\alpha_{em}|^{max}$ leads to additional limitations of $m/M$ and $q/e$ relations. From equality (16), we obtain $\alpha = \dfrac{m}{M_P}\dfrac{M}{M_P} = \dfrac{m}{M}\dfrac{M^2}{M_P^2}$;

$\dfrac{m}{M_P} = \sqrt{\dfrac{m}{M}}\sqrt{\alpha} = \dfrac{\alpha_Q}{\alpha_{em}}\sqrt{\alpha_{fs}}\dfrac{q}{e}$. The limiting value is $\rho_+ = \alpha\left(1 + \sqrt{1-k^2}\right) \sim |\alpha_{em}|^{max}$. Here, $k = \alpha_Q/\alpha$ and $0 < k < 1$. Then, $\left(\dfrac{m}{M}\right)_{lim} = \dfrac{k^2 \alpha_{fs}(q/e)^2}{\left(1+\sqrt{1-k^2}\right)^2}\dfrac{1}{\alpha}$.

### 5.2 Availability of two even horizons $\rho_+, \rho_-$; $\alpha^2 > \alpha_Q^2$. Domains of $\rho \in (0, \rho_-]$

In this case, there exists the $m_\varphi$-degenerate solution of $\varepsilon = \alpha_{em}/\rho_-$. For bound states, $-1 < \varepsilon < 1$, therefore, $-\rho_- < \alpha_{em} < \rho_-$. The solution of $\varepsilon = \alpha_{em}/\rho_-$ includes states with like and opposite charges of the RN field source and the fermion. The case of $\varepsilon = 0$, $\alpha_{em} = 0$ corresponds to an uncharged Dirac particle.

Fig. 4 presents integral curves of equation (70) in the neighborhood of the irregular singular point $\rho = \rho_-$. One can see separatrices (red curves) corresponding to asymptotics (78) - (80) with $A_1 = -4$ и $A_2 = -4/3$. One of the separatrices, beginning at $(\rho_-, 0)$, corresponds to the physically acceptable asymptotics with $A_1 = -4$. The integral curves of $\Phi(\rho - \rho_-)$ are $\Phi$-periodic with period $\pi$.

Fig. 5 presents integral curves of equation (70) in the neighborhood of the irregular singular point $\rho = 0$. One can see separatrices (red curves), one of which corresponds to physically acceptable asymptotics (81) - (84) with solution of equation (82) $B_1 = 2/3$. The integral curves of $\Phi(\rho)$ are $\Phi$-periodic with period $\pi$.

While integrating (70) from left to right (from $\rho = 10^{-8}$ to $\rho_- - \rho = 10^{-8}$), there was a correct choice of asymptotics $\Phi(\rho), P(\rho), \psi_F(\rho)$ in the calculations both in the neighborhood of $\rho = 0$ and in the neighborhood of $\rho = \rho_-$.

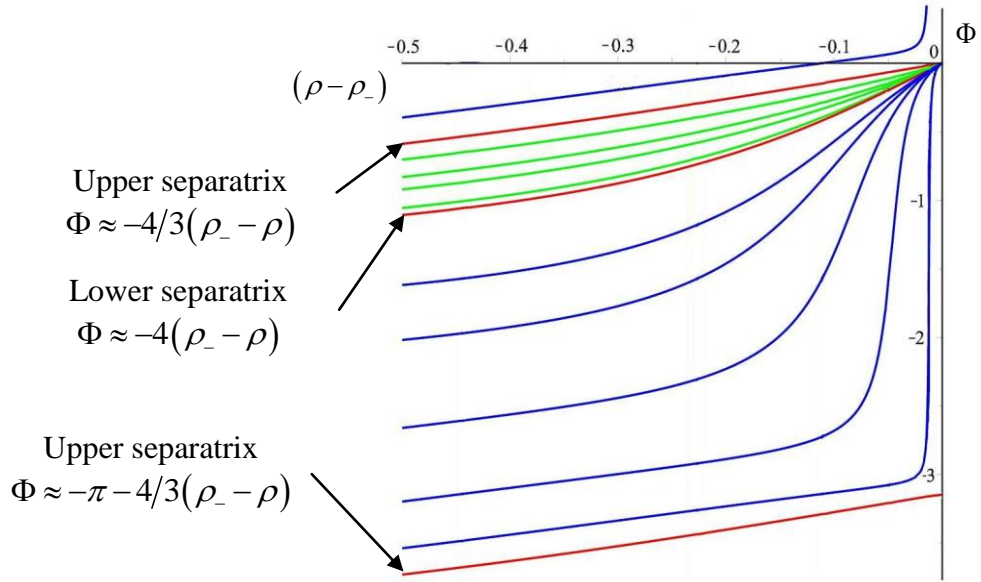

Fig.4. Integral curves of $\Phi(\rho - \rho_-)$ in the neighborhood of $\rho_-$.

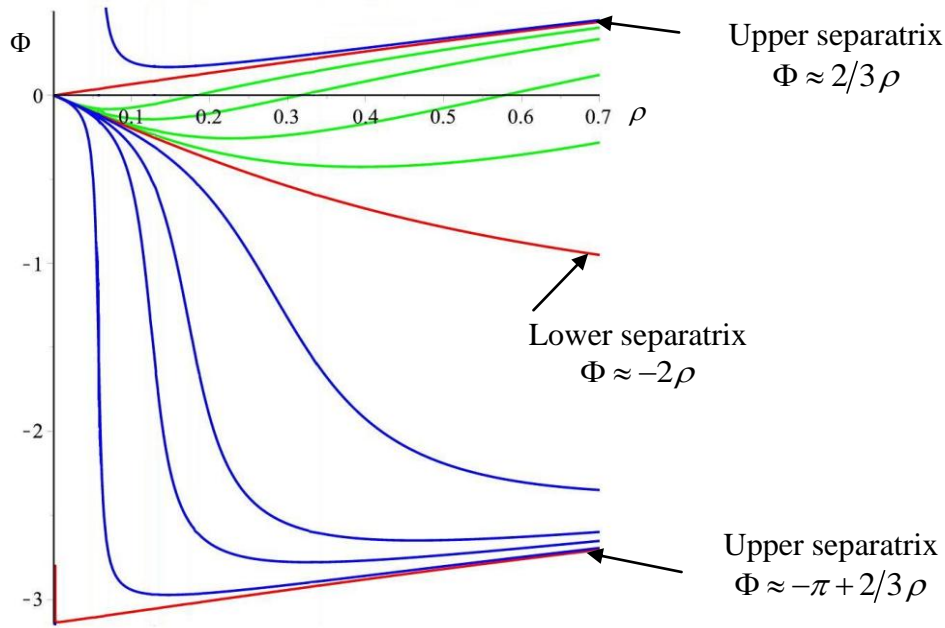

Fig. 5. Integral $\Phi(\rho)$ curves in the neighborhood of $\rho = 0$.

### 5.2.1 Physical acceptability range of solution $\varepsilon = \alpha_{em}/\rho_-$ and computed results

The solution of $\varepsilon = 0$ at $\alpha_{em} = 0$ is formally valid for any value of the gravitational coupling constant $\alpha$. We will restrict ourselves to consideration of less exotic systems with the event horizon $\rho_- \geq 1$, i.e. with the radii of the internal event horizon larger or comparable with the Compton wave length of fermion $l_c$. From inequality of $\rho_- = \alpha\left(1 - \sqrt{1-k^2}\right) \geq 1$, automatic restrictions by the value $\alpha$ and the ratio of $k = \alpha_Q/\alpha$ $(0 < k < 1)$ arise.

Table 4 presents distances $\rho_m$ from maxima of probability densities to the event horizon of $\rho_-$, obtained from the calculations, for solution of $\varepsilon = 0$, for $\kappa = 0.909$; $\alpha = 10$; $\rho_- = 5.832$ and for different $j, l$. Fig. 6 presents the normalization probability densities (85) and integral probabilities (86) for some values of $j, l$ and with the parameters of Table 4.

The range of values of the electromagnetic coupling constant $\alpha_{em}$ for the solution of $\varepsilon = \alpha_{em}/\rho_-$ is determined by the range of fermion energy for bound states of $-1 < \varepsilon < 1$. The maximal value of $|\alpha_{em}|^{max}$ is $\rho_- \sim |\alpha_{em}|^{max}$. Table 5 presents the values of $\rho_m$ versus variations in admissible values of $\alpha_{em}$ for $\alpha = 10$; $\alpha_Q = 9.09$ and for various $j, l$ values.

Table 4. The values of $\rho_m$ for $\varepsilon = 0$; $\kappa = 0.909$; $\alpha = 10$; $\rho_- = 5.832$ versus $\kappa$ (or $j, l$).

| $\kappa$ | $-1$ | $+1$ | $-2$ | $+2$ | $-3$ | $+3$ |
|---|---|---|---|---|---|---|
| $(j,l)$ | $(j=1/2, l=0)$ | $(j=1/2, l=1)$ | $(j=3/2, l=1)$ | $(j=3/2, l=2)$ | $(j=5/2, l=2)$ | $(j=5/2, l=3)$ |
| $\rho_m$ | 0.0152 | 0.0151 | 0.014 | 0.0138 | 0.0123 | 0.0121 |

Table 5. The values of $\rho_m$ for $\alpha = 10$; $\alpha_Q = 9.09$; $\rho_- = 5.832$ and various $\kappa$ versus admissible $\alpha_{em}$ values.

| $\kappa$ | \multicolumn{5}{c|}{$-1$} | \multicolumn{5}{c|}{$+1$} |
|---|---|---|---|---|---|---|---|---|---|---|
| $\alpha_{em}$ | 5.832 | 2.916 | 0 | $-2.916$ | $-5.832$ | 5.832 | 2.916 | 0 | $-2.916$ | $-5.832$ |
| $\varepsilon$ | $+1$ | $+0.5$ | 0 | $-0.5$ | $-1$ | $+1$ | $+0.5$ | 0 | $-0.5$ | $-1$ |
| $\rho_m$ | 0.0155 | 0.0154 | 0.0152 | 0.01514 | 0.0151 | 0.0149 | 0.015 | 0.0151 | 0.0152 | 0.0154 |
| $\kappa$ | \multicolumn{5}{c|}{$-2$} | \multicolumn{5}{c|}{$+2$} |
| $\alpha_{em}$ | 5.832 | 2.916 | 0 | $-2.916$ | $-5.832$ | 5.832 | 2.916 | 0 | $-2.916$ | $-5.832$ |
| $\varepsilon$ | $+1$ | $+0.5$ | 0 | $-0.5$ | $-1$ | $+1$ | $+0.5$ | 0 | $-0.5$ | $-1$ |
| $\rho_m$ | 0.0145 | 0.0142 | 0.014 | 0.0139 | 0.0137 | 0.0135 | 0.0137 | 0.0138 | 0.014 | 0.0142 |
| $\kappa$ | \multicolumn{5}{c|}{$-3$} | \multicolumn{5}{c|}{$+3$} |
| $\alpha_{em}$ | 5.832 | 2.916 | 0 | $-2.916$ | $-5.832$ | 5.832 | 2.916 | 0 | $-2.916$ | $-5.832$ |
| $\varepsilon$ | $+1$ | $+0.5$ | 0 | $-0.5$ | $-1$ | $+1$ | $+0.5$ | 0 | $-0.5$ | $-1$ |
| $\rho_m$ | 0.0129 | 0.0126 | 0.0123 | 0.0121 | 0.012 | 0.0117 | 0.0119 | 0.0121 | 0.0124 | 0.0126 |

The calculations show that the bound fermions are near the internal neighborhood of event horizon $\rho_-$, with the overwhelming probability. The localization of fermions increases with increase in $j, l$.

The data of Table 5 show that the maximal values of $\rho_m$ at $\alpha_{em} > 0$ for the states of $j = l + 1/2$ $(\kappa < 0)$ are achieved at the values of $\rho_- \sim \alpha_{em}^{\max}$. For opposite charges of the RN field source and the fermion $(\alpha_{em} < 0)$, the minimal values of $\rho_m$ are achieved at the minimally admissible value of $\alpha_{em}^{\min} \sim -\rho_-$. Inversely, the maximal values of $\rho_m$ for the states of $j = l - 1/2$ $(\kappa > 0)$ are achieved at $\alpha_{em}^{\min} \sim -\rho_-$, and the minimal values are achieved at $\rho_- \sim \alpha_{em}^{\max}$.

As well as in previous section 5.1, the solution of $\varepsilon = \alpha_{em}/\rho_-$ is possible only at definite ratios between masses and charges of the RN field source and the fermion. As before, we assume that $M \geq M_P$; $\alpha_Q = k\alpha$, where $0 < k < 1$, $m/M \leq 10^{-2}$, $\left(\dfrac{m}{M}\right)_{\lim} = \dfrac{k^2 \alpha_{fs} (q/e)^2}{\left(1 - \sqrt{1 - k^2}\right)^2} \dfrac{1}{\alpha}$. We must also use the limiting value of $\rho_- = 1$.

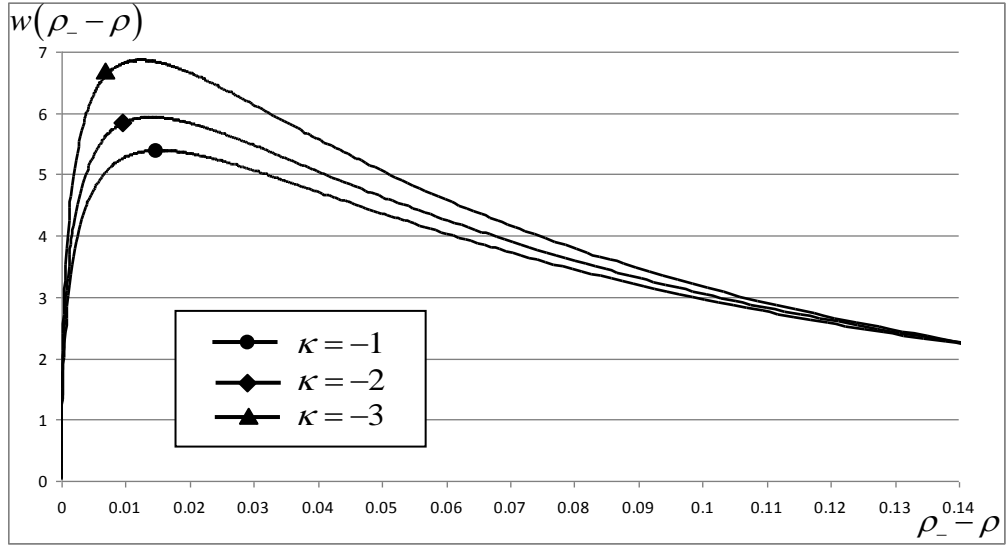

a)

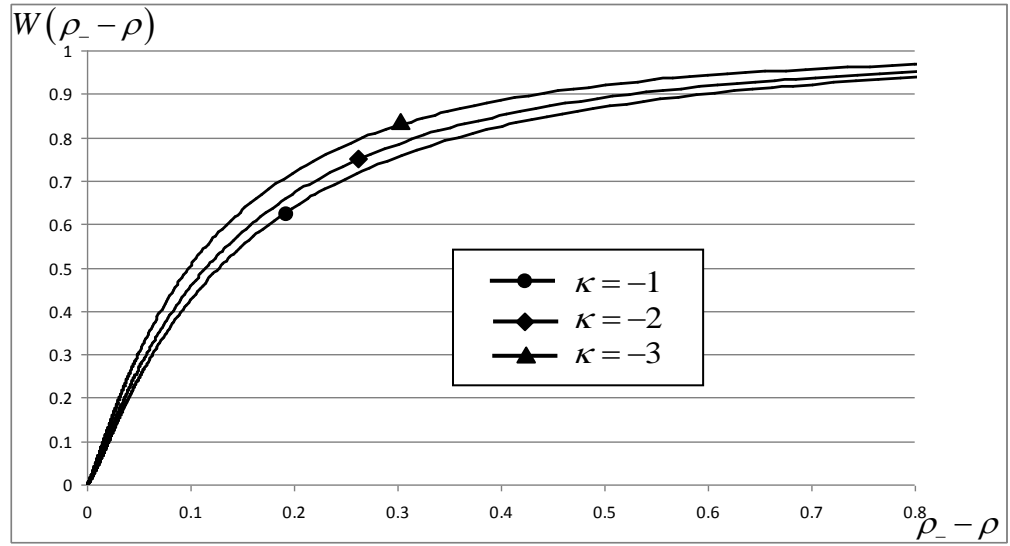

b)

Fig. 6. The curves: a) $w(\rho_- - \rho)$, b) $W(\rho_- - \rho)^3$ for bound states with $\varepsilon = 0$; $\alpha = 10;\ \alpha_Q = 0.909;\ \rho_- = 5.832;\ \kappa = -1, -2, -3$.

**5.3 Extreme Reissner-Nordström field $\left(\rho_+ = \rho_- = \alpha;\ \alpha^2 = \alpha_Q^2\right)$. Domain of wave functions:** $\rho \in (0, \alpha],\ \rho \in [\alpha, \infty)$

Earlier in section 3.1, we showed impossibility of existence of stationary bound states of fermions with $|\varepsilon| < 1$ both outside and inside the single event horizon $\rho_\pm = \alpha$. At $\varepsilon = \alpha_{em}/\alpha$, the impenetrable barrier of $\rho_{cl}$ is on the event horizon (see equality (57)).

For illustration, Fig.7 presents the characteristic form of potential $U_{eff}^F(\rho)$ for the extreme RN field. As it is noted in [5], existence of the bound state of the fermion with

---

[3] Curve $W(\rho)$ is calculated by integration from right to left from $\rho_- - \rho_{max} = 10^{-8}$.

$\varepsilon = \alpha_{em}/\alpha > 1$ is possible in the domain under the event horizon $\rho \in (0, \alpha]$ for like charges of the fermion and the RN field source. In this case, the value of $\varepsilon$ should be higher than the minimum of the potential $U_{eff}^F(\rho)$ in the domain under consideration.

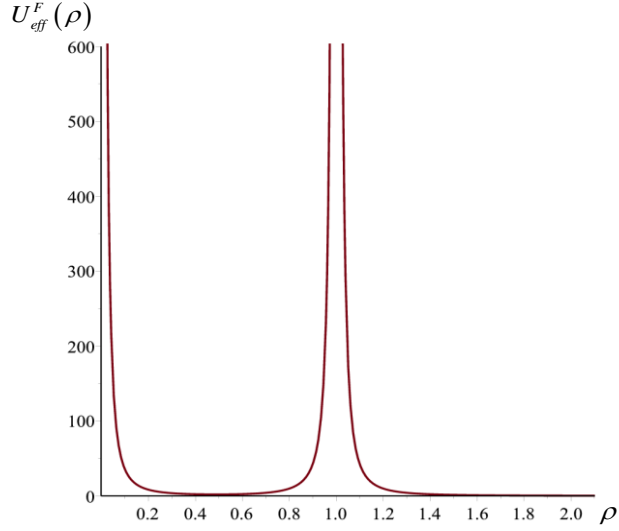

Fig.7. The curve of $U_{eff}^F(\rho)$ in the extreme RN field at
$\alpha = 1;\ \alpha_Q = 1;\ \alpha_{em} = 0.9;\ \kappa = -1;\ \varepsilon = 0.9;\ \rho_+ = \rho_- = 1$.

**5.4 Naked Reissner-Nordström singularity $(\alpha^2 < \alpha_Q^2)$. Domain of wave functions:** $\rho \in (0, \infty)$

As distinct from sections 5.1, 5.2, the $n, j, l$ - non-degenerate discrete energy spectrum exists for charged fermions in the field of RN naked singularity at definite values of physical parameters (see section 3.2). The discrete spectrum also exists in case of uncharged fermions.

For illustration, two variants of atomic systems were examined in numerical calculations.

The first variant:

$$\alpha = 0.025;\ |\alpha_Q|/\alpha = 2;\ \alpha_{em} = -0.1 \text{ or } \alpha_{em} = 0. \tag{91}$$

A particle with $M = 0.585 M_P$ and charge $+13.7e$ is the source of the RN field; a particle with $m = 0.043 M_P$ and with charge $-e$, or an uncharged particle with $m = 0.043 M_P$ is chosen as the fermion.

The second variant:

$$\alpha = 0.25;\ |\alpha_Q|/\alpha = 2;\ \alpha_{em} = -1 \text{ or } \alpha_{em} = 0. \tag{92}$$

A particle with $M = 5.85 M_P$ and charge $+137e$ is the source of the RN field; a particle with $m = 0.043 M_P$ and with charge $-e$, or an uncharged particle with $m = 0.043 M_P$ is chosen as the fermion.

To illustrate the existence of stationary bound states of fermions charged identically to the field source of the naked RN singularity at $\varepsilon > \alpha_{em}/\alpha$ (see item 3.2, Table 1), the second variant was calculated with $\alpha_{em} = 0.1$.

In numerical calculations, equation (70) was integrated from left to right (from $\rho_{min}$ to $\rho_{max}$). The good mathematical convergence of the results is ensured when $\rho_{min} = 10^{-8}$, $\rho_{max} = 10^{7}$ are chosen. As well as in section 5.2, physically acceptable asymptotics (81) - (84) are ensured in calculations in the neighborhood of $\rho = 0$ with solution of equation (82) $B_1 = 2/3$.

The energy levels were determined in the calculations at the points where the function $\Phi(\varepsilon, \rho_{max}) = \Phi(\varepsilon)|_{\rho \to \rho_{max}}$ varies stepwise by $\pi$, in compliance with (87).

Tables 6 - 8 present calculated values of the discrete spectrum $(1-\varepsilon_n)$ versus some values of quantum numbers $n, j, l$ for the said alternate values of $\alpha, |\alpha_Q|, \alpha_{em}$.

The computational values of $(1-\varepsilon_n)$ are determined with an accuracy to the first sign after the point. Numerical values in the second signs after the point can vary after special calculations performed for convergence of mathematical results. This paper presents discrete spectra for qualitative evaluations without carrying out precision calculations.

Table 6. Numerical values of $1-\varepsilon_n$ at opposite signs of $Q, e$ charges for two alternate values of $\alpha, \alpha_Q, \alpha_{em}$.

|  | $n=1, \kappa=-1$ $j=1/2, l=0$ | $n=2, \kappa=-1$ $j=1/2, l=0$ | $n=3, \kappa=-1$ $j=1/2, l=0$ | $n=2, \kappa=+1$ $j=1/2, l=1$ | $n=3, \kappa=+1$ $j=1/2, l=1$ |
|---|---|---|---|---|---|
| $\alpha=0.025; \alpha_Q=0.05; \alpha_{em}=-0.1$ | $7.85\cdot10^{-3}$ | $1.97\cdot10^{-3}$ | $8.74\cdot10^{-4}$ | $1.97\cdot10^{-3}$ | $8.73\cdot10^{-4}$ |
| $\alpha=0.25; \alpha_Q=0.5; \alpha_{em}=-1$ | $5.18\cdot10^{-1}$ | $1.98\cdot10^{-1}$ | $9.58\cdot10^{-2}$ | $3.21\cdot10^{-1}$ | $1.35\cdot10^{-1}$ |

|  | $n=2, \kappa=-2$ $j=3/2, l=1$ | $n=3, \kappa=-2$ $j=3/2, l=1$ | $n=3, \kappa=+2$ $j=3/2, l=2$ | $n=3, \kappa=-3$ $j=5/2, l=2$ |
|---|---|---|---|---|
| $\alpha=0.025; \alpha_Q=0.05; \alpha_{em}=-0.1$ | $1.96\cdot10^{-3}$ | $8.71\cdot10^{-4}$ | $8.7\cdot10^{-4}$ | $8.69\cdot10^{-4}$ |
| $\alpha=0.25; \alpha_Q=0.5; \alpha_{em}=-1$ | $2.22\cdot10^{-1}$ | $1.08\cdot10^{-1}$ | $1.1\cdot10^{-1}$ | $9.31\cdot10^{-2}$ |

Table 7. Numerical values of $1-\varepsilon_n$ for uncharged fermions for two alternate values of $\alpha, \alpha_Q$

|  | $n=1, \kappa=-1$ $j=1/2, l=0$ | $n=2, \kappa=-1$ $j=1/2, l=0$ | $n=3, \kappa=-1$ $j=1/2, l=0$ | $n=2, \kappa=+1$ $j=1/2, l=1$ | $n=3, \kappa=+1$ $j=1/2, l=1$ |
|---|---|---|---|---|---|
| $\alpha=0.025; \alpha_Q=0.05; \alpha_{em}=0$ | $3.11\cdot10^{-4}$ | $7.81\cdot10^{-5}$ | $3.47\cdot10^{-5}$ | $7.82\cdot10^{-5}$ | $3.47\cdot10^{-5}$ |
| $\alpha=0.25; \alpha_Q=0.5; \alpha_{em}=0$ | $2.31\cdot10^{-2}$ | $6.95\cdot10^{-3}$ | $3.24\cdot10^{-3}$ | $7.84\cdot10^{-3}$ | $3.52\cdot10^{-3}$ |

|  | $n=2, \kappa=-2$ $j=3/2, l=1$ | $n=3, \kappa=-2$ $j=3/2, l=1$ | $n=3, \kappa=+2$ $j=3/2, l=2$ | $n=3, \kappa=-3$ $j=5/2, l=2$ |
|---|---|---|---|---|
| $\alpha=0.025; \alpha_Q=0.05; \alpha_{em}=0$ | $7.81\cdot10^{-5}$ | $3.47\cdot10^{-5}$ | $3.47\cdot10^{-5}$ | $3.47\cdot10^{-5}$ |
| $\alpha=0.25; \alpha_Q=0.5; \alpha_{em}=0$ | $7.53\cdot10^{-3}$ | $3.43\cdot10^{-3}$ | $3.46\cdot10^{-3}$ | $3.42\cdot10^{-3}$ |

Table 8. Numerical values of $1-\varepsilon_n$ at like charges of $Q, e$.

|  | $n=1, \kappa=-1$ $j=1/2, l=0$ | $n=2, \kappa=-1$ $j=1/2, l=0$ | $n=3, \kappa=-1$ $j=1/2, l=0$ | $n=2, \kappa=+1$ $j=1/2, l=1$ | $n=3, \kappa=+1$ $j=1/2, l=1$ |
|---|---|---|---|---|---|
| $\alpha=0.25;\ \alpha_Q=0.5;\ \alpha_{em}=0.1$ | $8.37\cdot 10^{-3}$ | $2.47\cdot 10^{-3}$ | $1.15\cdot 10^{-3}$ | $2.74\cdot 10^{-3}$ | $1.24 10^{-3}$ |

|  | $n=2, \kappa=-2$ $j=3/2, l=1$ | $n=3, \kappa=-2$ $j=3/2, l=1$ | $n=3, \kappa=+2$ $j=3/2, l=2$ | $n=3, \kappa=-3$ $j=5/2, l=2$ |
|---|---|---|---|---|
| $\alpha=0.25;\ \alpha_Q=0.5;\ \alpha_{em}=0.1$ | $2.68\cdot 10^{-3}$ | $1.22\cdot 10^{-3}$ | $1.23\cdot 10^{-3}$ | $1.225\cdot 10^{-3}$ |

For illustration, Fig. 8 presents the curves of $\Phi(1-\varepsilon, \rho_{max})$ for the second variant with $\kappa=-1$.

Figures 9, 10 present normalized densities of probability for eigenfunctions of states $1S_{1/2}$ $(n=1, \kappa=-1, l=0, j=1/2)$ and $2P_{1/2}$ $(n=2, \kappa=1, l=1, j=1/2)$. The curves were obtained for the two variants of atomic systems examined above. For all the variants, both charged and uncharged Dirac particles were considered. The wave functions and probability densities for eigenvalues of $\varepsilon_n$ were determined by integration of equations (70), (71) from right to left using boundary condition (72). In this case, it is seen from Fig.5 that it is impossible to achieve a physically acceptable separatrix with asymptotics (81), with solution of equation (82) $B_1 = 2/3$ in numerical calculations in the neighborhood of $\rho = 0$. In our calculations, asymptotics (81) - (84) with $B_1 = 2/3$ were joined with the appropriate functions from numerical calculations at $\rho = 10^{-2} \div 10^{-3}$. Such a procedure is actually not reflected on the curves of Figs. 9, 10.

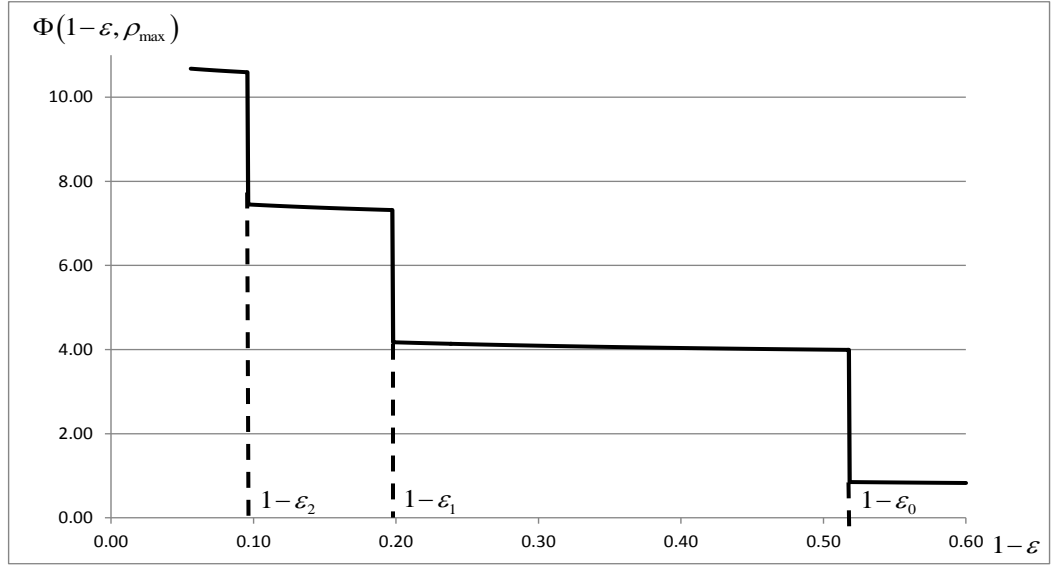

Fig.8. The relations of $\Phi(1-\varepsilon, \rho_{max})$ in the calculations with $\alpha=0.25;\ \alpha_Q=0.5;$ $\alpha_{em}=-1;\ \kappa=-1;\ \rho_{max}=10^7$.

In Fig. 8, $\varepsilon_0, \varepsilon_1, \varepsilon_2$ correspond to energies of bound states of $1S_{1/2}, 2S_{1/2}, 3S_{1/2}$. For the rest of the energies, $\Phi(1-\varepsilon, \rho_{\max}) \sim \text{arctg}\, \dfrac{1}{\sqrt{1-\varepsilon^2}} + k\pi$ in compliance with (73).

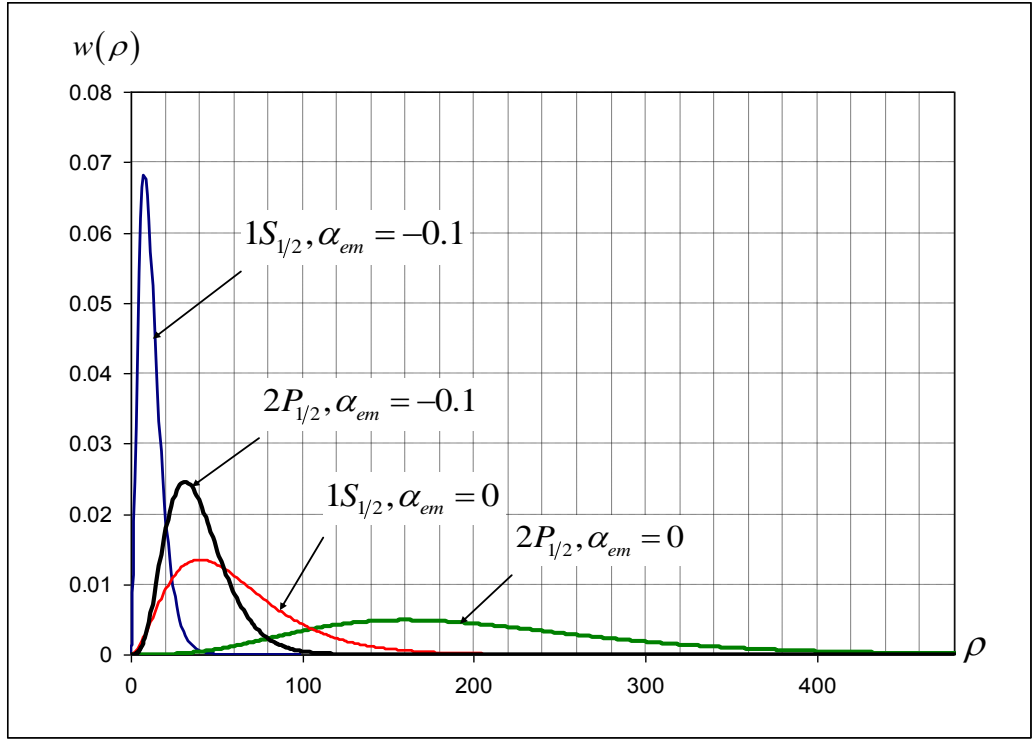

Fig. 9. Normalized probability density for $1S_{1/2}, 2P_{1/2}$-states of charged and uncharged Dirac particles at $\alpha = 0.025$, $\alpha_Q = 0.05$.

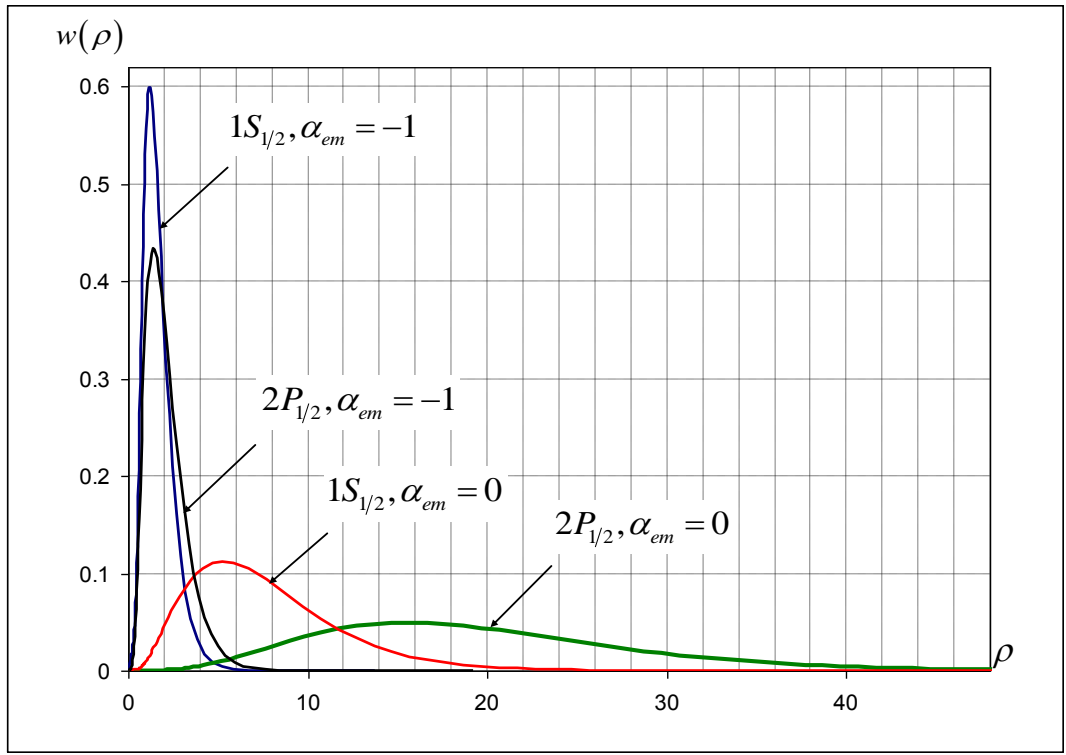

Fig. 10. Normalized probability density for $1S_{1/2}, 2P_{1/2}$-states of charged and uncharged Dirac particles at $\alpha = 0.25$, $\alpha_Q = 0.5$.

Based on the computed results, let us note the existence of stationary bound states of uncharged Dirac particles arising due to gravitational interaction alone. In general, at low values of $\alpha, \alpha_Q, \alpha_{em}$ (the first variant), discrete spectra are similar to spectra of hydrogen-like atoms with weak dependence of binding energy of $(1-\varepsilon_n)$ on values of orbital and total angular momenta of particles. In the second variant, rather a strong coupling of Dirac particles with the field of naked RN singularity is ensured with stable removal of degeneration of energy levels with the given value of $n$ and with $\pm\kappa$ values.

In general, the behavior of probability densities is of the same character as that when considering atomic system in the Minkowsky space. It can be noticed that for $1S_{1/2}$-states, the Bohr radius $r_B = \hbar^2/me^2$ (in dimensionless units of $\rho_B = (\alpha_{em})^{-1}$) is close to appropriate maxima of probability density

Variant 1: $\rho_B = 10$; $\rho_{w_{max}} = 7.7$ - Fig. 9;

Variant 2: $\rho_B = 1$; $\rho_{w_{max}} = 1.18$ - Fig. 10.

The maxima of the probability density are displaced towards larger values of $\rho$ in all the variants for uncharged Dirac particles (purely gravitational interaction).

For $2P_{1/2}$ - states, the appropriate maxima of probability densities, as compared with $1S_{1/2}$ - states, are within the domain of larger values $\rho$.

## 6. Cosmic censorship

The hypothesis of cosmic censorship, proposed in [25], bans existence of singularities, not covered with event horizons, in nature. However, it is no complete proof of this hypothesis so far. Many of the researchers, along with black holes, consider generation of naked singularities, their stability and distinctive featuresin the course of experimental observations [26] - [31].

Geodesic incompleteness of naked RN singularity is well known for classical particles. Massive test particles, moving geodesically, do not achieve singularity due to the repulsive nature of inner regions of the RN metric.

It is shown in [6] that static metrics exist with time-like singularities (Schwarzschild and Reissner-Nordström metrics are among them), which prove to be completely non-singular when considering quantum mechanics of spinless particles. The authors [6] note that Schwarzschild and Reissner-Nordström solutions for spinless particles become singular only at negative $M < 0$. In the classical case for $M < 0$, both the solutions are geodesically complete. In our paper, the results were validated [6] as applied to motion of fermions in the field of naked RN singularity.

Indeed, the leading singularity of the effective potential $U_{eff}^F$ in the neighborhood of singularity represents an infinitely large potential barrier (41)

$$U_{eff}^F \big|_{\rho \to 0} = \frac{3}{8\rho^2} + O\left(\frac{1}{\rho}\right).$$

Since the naked RN singularity in quantum mechanics is covered with repulsive barrier (41), the presence of this singularity imposes no threat to cosmic censorship.

### 7. Degenerate states and particles of dark matter

In presence of event horizons $\rho_+, \rho_-$, degenerate stationary bound states of fermions in the RN field determined by solutions of $\varepsilon = \alpha_{em}/\rho_\pm$, can be candidates for the role of dark matter particles.

Indeed, let us consider the solution of $\varepsilon = \alpha_{em}/\rho_-$, for instance. If an atomic system emerges during generation of a RN collapsar, with bound fermions being near the internal neighborhood of the event horizon $\rho_-$, and if the source charge of the RN field in this case is compensated by a total charge of bound fermions, then for the outer world, such an atomic system interacts with other objects only gravitationally. Due to absence of quantum transitions between the states with different $\kappa$ (or $j,l$), such a system neither emits nor absorbs light and other types of emission. Such a system can be discovered only through gravitational interaction.

The system of bound fermions in the RN field with the energy of $\varepsilon = \alpha_{em}/\rho_+$ can be considered as the second atomic system. In this case, the fermions are, with the overwhelming probability, near the external neighborhood $\rho_+$ and such an atomic system interacts with other external objects only gravitationally, provided the charge of the RN field source is compensated by the total charge of bound fermions. As well as in the first case, the atomic system neither emits nor absorbs light and other types of emission. It is possible to detect a charge of the RN field source in the given atomic system only by "knocking-out" part of fermions from their orbits due to external action.

Other atomic systems can be candidates for a particle of dark matter, with the energy of bound fermions partially with $\varepsilon = \alpha_{em}/\rho_+$, partially with $\varepsilon = \alpha_{em}/\rho_-$. The masses of such systems should be selected based on the condition of the best agreement with the Standard cosmological model.

It should be noted that atomic systems were considered as a zero approximation without consideration of interaction of bound fermions with each other and within the framework of applicability of single-particle quantum mechanics.

## 8. Conclusions

The consideration of stationary solutions of the self-conjugate second-order equation with the spinor wave function in quantum mechanics of fermion motion in the classical Reissner-Nordström field resulted in the following:

1. In presence of two event horizons $\rho_+, \rho_-$, there exist regular solutions with energies of $\varepsilon = \alpha_{em}/\rho_+$, $\varepsilon = \alpha_{em}/\rho_-$. These solutions represent degenerate stationary bound states of charged fermions with square integrable wave functions and with the domains of $\rho \in [\rho_+, \infty)$, $\rho \in (0, \rho_-]$. The wave functions weakly depend on $j, l$ and vanish on event horizons. The fermions in bound states are near the event horizons, with the overwhelming probability. The maxima of probability densities are away from event horizons at the distance of several fractions of Compton wave length of fermions.

2. Absence of stationary bound states of fermions with $|\varepsilon| < 1$ was proved for the extreme RN field with the single event horizon $\rho_+ = \rho_- = \alpha$.

3. For naked RN singularity $(\alpha^2 < \alpha_Q^2)$, the analysis of effective potentials and direct numerical solutions of the self-conjugate second-order equation at definite values of physical parameters showed existence of stationary bound states of charged fermions of the RN field. Bound states also exist for electrically uncharged fermions; these states are implemented only due to gravitational interaction forces.

4. Naked RN singularity is separated by infinitely high potential barrier of $\sim 3/8\rho^2$ for any quantum mechanical half-spin particle, irrespective of availability and sign of the electrical charge. This agrees with the conclusions of [3] with regard to motion of spinless particles. Presence of the repulsive barrier, covering the singularity, poses no threat to cosmic censorship.

Regular stationary solutions are absent for the Dirac equation with the bispinor wave function in the Reissner-Nordström field for the following reasons:

- in presence of event horizons, due to square nonintegrability of wave functions (see also [2]);

- for the extreme RN field in the domain outside the single event horizon because of non-fulfillment of one of the conditions of existence of bound states, proved in [3];

- for the extreme RN field in the domain under the event horizon and for the case of naked RN singularity due to existence of two solutions with square integrable wave functions in the neighborhood of the origin of coordinates $\rho = 0$ [4].

Electrically neutral atomic systems with the definite number of fermions in degenerate bound states with $\varepsilon = \alpha_{em}/\rho_+$, $\varepsilon = \alpha_{em}/\rho_-$ can be considered in the standard cosmological model as particles of dark matter. Atomic systems of such a type neither absorb nor emit light and other kinds of emissions and interact with the environment gravitationally.

So, our examination shows that use of the second-order equation (30) expands the opportunities of obtaining regular solutions of quantum mechanics of half-spin particle motion in external gravitational fields. .

## Acknowledgements

The authors express their gratitude to M. V. Gorbatenko and E.Yu. Popov for fruitful discussions and to A.L. Novoselova for the essential technical assistance in preparation of the paper.

### APPENDIX A
### Self-conjugate second-order equations for spinor wave functions of fermions in Schwarzschild and Reissner-Nordström fields

Let us introduce dimensionless variables and denotations in Hamiltonian 1

$$H_\eta = H_1 + V(\rho), \text{ where } V(\rho) = \alpha_{em}/\rho. \tag{A.1}$$

Taking into account (10) and (A.1), equation (5) has the form of

$$(\varepsilon - V(\rho) - H_1)\Psi_\eta(\rho,\theta,\varphi) = 0. \tag{A.2}$$

Let us multiply equation (A.2) on the left by operator $(\varepsilon - V(\rho) + H_1)$. Then,

$$(\varepsilon - V(\rho) + H_1)(\varepsilon - V(\rho) - H_1)\Psi_\eta(\rho,\theta,\varphi) = 0. \tag{A.3}$$

$$\begin{aligned}
&\left\{(\varepsilon-V)^2 - f_{R-N} + \left(f_{R-N}\frac{\partial}{\partial\rho} + \frac{1}{\rho} - \frac{\alpha}{\rho^2}\right)\left(f_{R-N}\frac{\partial}{\partial\rho} + \frac{1}{\rho} - \frac{\alpha}{\rho^2}\right) + \right. \\
&+ \frac{f_{R-N}}{\rho^2}\left[\left(\frac{\partial}{\partial\theta} + \frac{1}{2}\operatorname{ctg}\theta\right)\left(\frac{\partial}{\partial\theta} + \frac{1}{2}\operatorname{ctg}\theta\right) + \frac{1}{\sin^2\theta}\frac{\partial^2}{\partial\varphi^2} + i\Sigma^3\frac{\partial}{\partial\theta}\left(\frac{1}{\sin\theta}\right)\frac{\partial}{\partial\varphi}\right] + \\
&+ i\gamma^0\gamma^3 f_{R-N}\frac{dV}{d\rho} - i\gamma^0\gamma^3 f_{R-N}\frac{d}{d\rho}\left(\sqrt{f_{R-N}}\right) + \\
&\left. + f_{R-N}\frac{d}{d\rho}\left(\sqrt{f_{R-N}}\frac{1}{\rho}\right)\left[i\Sigma^2\left(\frac{\partial}{\partial\theta} + \frac{1}{2}\operatorname{ctg}\theta\right) - i\Sigma^1\frac{1}{\sin\theta}\frac{\partial}{\partial\varphi}\right]\right\}\Psi_\eta(\rho,\theta,\varphi) = 0.
\end{aligned} \tag{A.4}$$

As earlier, in (A.4), the equivalent substitution of matrices (12) was performed, $\Sigma^k = \begin{pmatrix} \sigma^k & 0 \\ 0 & \sigma^k \end{pmatrix}$.

The Dirac equations for upper and lower bispinor components

$$\Psi_\eta(\rho,\theta,\varphi,t) = \begin{pmatrix} U(\rho,\theta,\varphi) \\ W(\rho,\theta,\varphi) \end{pmatrix} e^{-i\varepsilon t} \qquad (A.5)$$

have the form of

$$\left(\varepsilon - V - \sqrt{f_{R-N}}\right)U = \left(-i\sigma^3\left(f_{R-N}\frac{\partial}{\partial\rho} + \frac{1}{\rho} - \frac{\alpha}{\rho^2}\right) - i\sigma^1\sqrt{f_{R-N}}\frac{1}{\rho}\left(\frac{\partial}{\partial\theta} + \frac{1}{2}\operatorname{ctg}\theta\right) - \right.$$

$$\left. -i\sigma^2\sqrt{f_{R-N}}\frac{1}{\rho\sin\theta}\frac{\partial}{\partial\varphi}\right)W,$$

$$\left(\varepsilon - V + \sqrt{f_{R-N}}\right)W = \left(-i\sigma^3\left(f_{R-N}\frac{\partial}{\partial\rho} + \frac{1}{\rho} - \frac{\alpha}{\rho^2}\right) - i\sigma^1\sqrt{f_{R-N}}\frac{1}{\rho}\left(\frac{\partial}{\partial\theta} + \frac{1}{2}\operatorname{ctg}\theta\right) - \right.$$

$$\left. -i\sigma^2\sqrt{f_{R-N}}\frac{1}{\rho\sin\theta}\frac{\partial}{\partial\varphi}\right)U.$$

(A.6)

As a result, equation (A.4), taking into account (A.6), can be written for one of the spinors $U(\rho,\theta,\varphi)$ or $W(\rho,\theta,\varphi)$. For the spinor $U(\rho,\theta,\varphi)$, equation (A.4) has the form of

$$\left\{(\varepsilon - V)^2 - f_{R-N} + \left(f_{R-N}\frac{\partial}{\partial\rho} + \frac{1}{\rho} - \frac{\alpha}{\rho^2}\right)^2 + \right.$$

$$+ \frac{f_{R-N}}{\rho^2}\left[\left(\frac{\partial}{\partial\theta} + \frac{1}{2}\operatorname{ctg}\theta\right)^2 + \frac{1}{\sin^2\theta}\frac{\partial^2}{\partial\varphi^2} + i\sigma^3\frac{\partial}{\partial\theta}\left(\frac{1}{\sin\theta}\right)\frac{\partial}{\partial\varphi}\right] +$$

$$+ f_{R-N}\frac{d}{d\rho}\left(\sqrt{f_{R-N}}\frac{1}{\rho}\right)\left[i\sigma^2\left(\frac{\partial}{\partial\theta} + \frac{1}{2}\operatorname{ctg}\theta\right) - i\sigma^1\frac{1}{\sin\theta}\frac{\partial}{\partial\varphi}\right] +$$

$$+ \left(f_{R-N}\frac{d}{d\rho}\left(\sqrt{f_{R-N}}\right) - f_{R-N}\frac{dV}{d\rho}\right)\frac{1}{\varepsilon - V + \sqrt{f_{R-N}}}\left[-f_{R-N}\frac{\partial}{\partial\rho} - \frac{1}{\rho} + \frac{\alpha}{\rho^2} - \right.$$

$$\left.\left. -i\sigma^2\sqrt{f_{R-N}}\frac{1}{\rho}\left(\frac{\partial}{\partial\theta} + \frac{1}{2}\operatorname{ctg}\theta\right) + i\sigma^1\sqrt{f_{R-N}}\frac{1}{\rho\sin\theta}\frac{\partial}{\partial\varphi}\right]\right\}U(\rho,\theta,\varphi) = 0.$$

(A.7)

Then, the variables can be separated. It follows from representation (10) that

$$U(r,\theta,\varphi) = F(\rho)\xi(\theta)e^{im_\varphi\varphi}. \qquad (A.8)$$

Using Brill-Wheeler equation (11) and its squared representation [16]

$$\left[\left(\frac{\partial}{\partial\theta} + \frac{1}{2}\operatorname{ctg}\theta\right)^2 + \frac{1}{\sin^2\theta}\frac{\partial^2}{\partial\varphi^2} + i\sigma^3\frac{\partial}{\partial\theta}\left(\frac{1}{\sin\theta}\right)\frac{\partial}{\partial\varphi}\right]\xi(\theta)e^{im_\varphi\varphi} = -\kappa^2\xi(\theta)e^{im_\varphi\varphi}, \qquad (A.9)$$

we can obtain the second-order equation for the radial function $F(\rho)$

$$\left\{(\varepsilon-V)^2 - f_{R-N} + \left(f_{R-N}\frac{\partial}{\partial\rho} + \frac{1}{\rho} - \frac{\alpha}{\rho^2}\right)^2 - \frac{f_{R-N}\kappa^2}{\rho^2} + f_{R-N}\kappa\frac{d}{d\rho}\left(\sqrt{f_{R-N}}\frac{1}{\rho}\right) - \right.$$

$$-\left(f_{R-N}\frac{d}{d\rho}\left(\sqrt{f_{R-N}}\right) - f_{R-N}\frac{dV}{d\rho}\right)\frac{1}{\varepsilon - V + \sqrt{f_{R-N}}}\frac{\kappa\sqrt{f_{R-N}}}{\rho} - \quad \text{(A.10)}$$

$$\left.-\left(f_{R-N}\frac{d}{d\rho}\left(\sqrt{f_{R-N}}\right) - f_{R-N}\frac{dV}{d\rho}\right)\frac{1}{\varepsilon - V + \sqrt{f_{R-N}}}\left(f_{R-N}\frac{\partial}{\partial\rho} + \frac{1}{\rho} - \frac{\alpha}{\rho^2}\right)\right\}F(\rho) = 0.$$

The third and the last summands in equation (A.10) are not self-conjugate. Let us perform non-unitary similarity transformation for self-conjugacy of (A.10).

$$F(\rho) = g_F^{-1}(\rho)\psi_F(\rho). \quad \text{(A.11)}$$

If we denote in equation (15) that

$$A(\rho) = -\frac{1}{f_{R-N}}\left(\frac{1 + \kappa\sqrt{f_{R-N}}}{\rho} - \frac{\alpha}{\rho^2}\right), \quad \text{(A.12)}$$

$$B(\rho) = \frac{1}{f_{R-N}}\left(\varepsilon - \frac{\alpha_{em}}{\rho} + \sqrt{f_{R-N}}\right), \quad \text{(A.13)}$$

$$C(\rho) = -\frac{1}{f_{R-N}}\left(\varepsilon - \frac{\alpha_{em}}{\rho} + \sqrt{f_{R-N}}\right), \quad \text{(A.14)}$$

$$D(\rho) = -\frac{1}{f_{R-N}}\left(\frac{1 - \kappa\sqrt{f_{R-N}}}{\rho} - \frac{\alpha}{\rho^2}\right) \quad \text{(A.15)}$$

and besides introduce the denotations of

$$A_F(\rho) = -\frac{1}{B}\frac{dB}{d\rho} - A - D, \quad \text{(A.16)}$$

$$A_G(\rho) = -\frac{1}{C}\frac{dC}{d\rho} - A - D, \quad \text{(A.17)}$$

then, the sought transformation is

$$g_F(\rho) = \exp\frac{1}{2}\int A_F(\rho')d\rho'. \quad \text{(A.18)}$$

As the result, we write equation (A.10) as

$$\hat{M}F(\rho) = 0,$$

then, the transformed self-conjugate equation has the view of

$$g_F\hat{M}g_F^{-1}\psi_F(\rho) = 0 \quad \text{(A.19)}$$

Equation (A.19) can be written in the form of the Schrödinger-type second-order equation with the effective potential $U_{eff}^F(\rho)$

$$\frac{d^2\psi_F}{d\rho^2} + 2\left(E_{Schr} - U_{eff}^F\right)\psi_F = 0, \qquad (A.20)$$

where

$$E_{Schr} = \frac{1}{2}\left(\varepsilon^2 - 1\right), \qquad (A.21)$$

$$U_{eff}^F = -\frac{1}{4}\frac{1}{B}\frac{d^2B}{d\rho^2} + \frac{3}{8}\left(\frac{1}{B}\frac{dB}{d\rho}\right)^2 - \frac{1}{4}(A-D)\frac{1}{B}\frac{dB}{d\rho} + \frac{1}{4}\frac{d}{d\rho}(A-D) +$$
$$+ \frac{1}{8}(A-D)^2 + \frac{1}{2}BC + E_{Schr}. \qquad (A.22)$$

The summand $E_{Schr}$ (A.21) in equation (A.20) is separated and simultaneously added to (A.22). This is done, on the one hand, for equation (A.20) to take the form of Schrödinger-type equation, on the other hand, to ensure the classical asymptotics of the effective potential at $\rho \to \infty$.

For the lower spinor $W(\rho,\theta,\varphi)$ with radial function $G(\rho)$, the appropriate formulas have the view

$$G(\rho) = g_G^{-1}\psi_G(\rho), \qquad (A.23)$$

$$g_G(\rho) = \exp\frac{1}{2}\int A_G(\rho')d\rho', \qquad (A.24)$$

$$\frac{d^2\psi_G}{d\rho^2} + 2\left(E_{Schr} - U_{eff}^G\right)\psi_G = 0, \qquad (A.25)$$

$$U_{eff}^G = -\frac{1}{4}\frac{1}{C}\frac{d^2C}{d\rho^2} + \frac{3}{8}\left(\frac{1}{C}\frac{dC}{d\rho}\right)^2 + \frac{1}{4}\frac{(A-D)}{C}\frac{dC}{d\rho} - \frac{1}{4}\frac{d}{d\rho}(A-D) +$$
$$+ \frac{1}{8}(A-D)^2 + \frac{1}{2}BC + E_{Schr}. \qquad (A.26)$$

# APPENDIX B

## Effective potential of the RN field in Schrödinger-type equation

According to (A.12) - (A.15), (A.22), we can obtain

$$\frac{3}{8}\frac{1}{B^2}\left(\frac{dB}{d\rho}\right)^2 = \frac{3}{8}\left\{\frac{f_{R-N}}{\omega+\sqrt{f_{R-N}}}\left[-\frac{1}{f_{R-N}^2}f'_{R-N}\left(\omega+\sqrt{f_{R-N}}\right)+\frac{1}{f_{R-N}}\left(\omega'+\frac{f'_{R-N}}{2\sqrt{f_{R-N}}}\right)\right]\right\}^2, \quad (B.1)$$

$$-\frac{1}{4}\frac{1}{B}\frac{d^2B}{d\rho^2} = -\frac{1}{4}\frac{f_{R-N}}{\omega+\sqrt{f_{R-N}}}\left[\frac{2}{f_{R-N}^3}(f'_{R-N})^2\left(\omega+\sqrt{f_{R-N}}\right)-\frac{1}{f_{R-N}^2}f''_{R-N}\left(\omega+\sqrt{f_{R-N}}\right)-\right.$$
$$\left.-\frac{2}{f_{R-N}^2}f'_{R-N}\left(\omega'+\frac{f'_{R-N}}{2\sqrt{f_{R-N}}}\right)+\frac{1}{f_{R-N}}\left(\omega''+\frac{f''_{R-N}}{2\sqrt{f_{R-N}}}-\frac{(f'_{R-N})^2}{4f_{R-N}^{3/2}}\right)\right], \quad (B.2)$$

$$\frac{1}{4}\frac{d}{d\rho}(A-D) = \frac{\kappa}{2}\left[\frac{1}{2}\frac{f'_{R-N}}{\rho f_{R-N}^{3/2}}+\frac{1}{\rho^2 f_{R-N}^{1/2}}\right], \quad (B.3)$$

$$-\frac{1}{4}\frac{(A-D)}{B}\frac{dB}{d\rho} = \frac{\kappa}{2\rho f_{R-N}^{1/2}}\left(-\frac{f'_{R-N}}{f_{R-N}}+\frac{1}{\omega+\sqrt{f_{R-N}}}\left(\omega'+\frac{f'_{R-N}}{2\sqrt{f_{R-N}}}\right)\right), \quad (B.4)$$

$$\frac{1}{8}(A-D)^2 = \frac{\kappa^2}{2f_{R-N}\rho^2}, \quad (B.5)$$

$$\frac{1}{2}BC = -\frac{1}{2f_{R-N}^2}\left(\omega^2-f_{R-N}\right). \quad (B.6)$$

In (B.1) - (B.6), $f_{R-N} = 1 - \frac{2\alpha}{\rho} + \frac{\alpha_Q^2}{\rho^2}$; $f'_{R-N} \equiv \frac{df_{R-N}}{d\rho} = \frac{2\alpha}{\rho^2} - \frac{2\alpha_Q^2}{\rho^3}$;

$f''_{R-N} \equiv \frac{df_{R-N}^2}{d\rho^2} = -\frac{4\alpha}{\rho^3} + \frac{6\alpha_Q^2}{\rho^4}$; $\omega = \varepsilon - \frac{\alpha_{em}}{\rho}$, $\omega' \equiv \frac{d\omega}{d\rho} = \frac{\alpha_{em}}{\rho^2}$, $\omega'' \equiv \frac{d^2\omega}{d\rho^2} = -\frac{2\alpha_{em}}{\rho^3}$.

The sum of expressions $E_{Schr}$ and (B.1) - (B.6) leads to the expression for the effective potential $U_{eff}^F$ (A.22).

## References

[1] V.P.Neznamov, I.I. Safronov. ZhETF, Vol. 154, No. 4 (10), 761-773 (2018).

[2] F. Finster, J. Smoller, and S.-T. Yau, J. Math. Phys., **41**, 2173 (2000).

[3] H. Schmid. Mathematische Nachrichten, **274-275**, 117 (2004); arxiv: math-ph/0207039v2.

[4] C. L. Pekeris and K. Frankowski, Proc. Natl. Acad. Sci. USA **83**, 1978 (1986).

[5] M. V. Gorbatenko, V. P. Neznamov, and E. Yu. Popov, Grav. Cosmol. **23**, 245 (2017), DOI: 10.1134/S0202289317030057; J. Phys.: Conf. Ser. **678** 012037 (2016), DOI:10.1088/1742-6596/678/1/012037; arxiv: 1511.05058 [gr-qc].

[6] G. T. Horowitz and D. Marolf, Phys. Rev. D **52**, 5670 (1995).

[7] H. Pruefer. Math. Ann. **95**, 499 (1926).

[8] I. Ulehla and M. Havlíček, Appl. Math. **25**, 358 (1980).

[9] I. Ulehla, M. Havlíček and J. Hořejší, Phys. Lett. A**82**, 64 (1981).

[10] I. Ulehla, Rutherford Laboratory preprint RL-82-095 (1982).

[11] M. V. Gorbatenko and V. P. Neznamov. Phys. Rev. D**82**, 104056 (2010); arxiv:1007.4631 [gr-qc].

[12] M. V. Gorbatenko and V. P. Neznamov. Phys. Rev. D, **83**, 105002 (2011); arxiv:1102.4067v1 [gr-qc].

[13] M. V. Gorbatenko and V. P. Neznamov. J. Mod. Phys. **6**, 303 (2015); arxiv:1107.0844 [gr-qc].

[14] J. Schwinger. Phys. Rev. **130**, 800 (1963).

[15] D. R. Brill and J. A. Wheeler. Rev. Mod. Phys. **29**, 465 (1957).

[16] S. R. Dolan. Trinity Hall and Astrophysics Group, Cavendish Laboratory. Dissertation (2006).

[17] A. Lasenby, C. Doran, J. Pritchard, A. Caceres, and S. Dolan. Phys. Rev. D **72**, 105014 (2005).

[18] S. Dolan and D. Dempsey. Class. Quant. Grav. **32**, 184001 (2015).

[19] L. L. Foldy and S. A. Wouthuysen. Phys. Rev. **78**, 29 (1950); V. P. Neznamov Part. Nucl., **37** (1), 86 (2006); V. P. Neznamov, A. J. Silenko. J. Math. Phys. **50**, 122301 (2009).

[20] L. D. Landau and E. M. Lifshitz. *Quantum Mechanics. Nonrelativistic Theory*, Pergamon Press, Oxford (1965).

[21] J. Dittrich and P. Exner, J. Math. Phys. **26**, 2000 (1985).

[22] E. Hairer and G. Wanner. *Solving ordinary differential equations II. Stiff and Differential-Algebraic Problems.* Springer-Verlag (1991, 1996).

[23] W. Pieper and W. Griener. Zs. Phys. **218**, 327 (1969).

[24] M.-T. Wang, arxiv: 1605.04968.

[25] R. Penrose, Rivista del Nuovo Cimento, Serie I, **1**, Numero Speciale: 252 (1969).

[26] R. S. Virbhadra, D. Narasimba, and S. M. Chitre, Astron. Astrophys. **337**, 1 (1998).

[27] K. S. Virbhadra and G. F. R. Ellis, Phys. Rev. D **65**, 103004 (2002).

[28] K. S. Virbhadra and C. R. Keeton, Phys. Rev. D **77**, 124014 (2008).

[29] D. Dey, K. Bhattacharya and N.Sarkar, Phys. Rev. D **88**, 083532 (2013).

[30] P. S. Joshi, D. Malafaxina and Maragan, Class. Quant. Grav. **31**, 015002 (2014).

[31] A. Goel, R. Maity, P. Roy, and Tsarkar, Phys. Rev. D **91**, 104029 (2015); arxiv: 1504.01302 [gr-qc].